\documentclass[12pt]{article}
\usepackage{latexsym}
\usepackage{amssymb}
\usepackage{amsmath}
\usepackage{graphicx}

\def\be{\begin{equation}}
\def\ee{\end{equation}}
\def\bear{\begin{eqnarray}}
\def\eear{\end{eqnarray}}
\def\nn{\nonumber}

\newcommand\bra[1]{{\langle {#1}|}}
\newcommand\ket[1]{{|{#1}\rangle}}

\def\a{\alpha}

\def\d{\delta}

\def\r{\rho}
\def\t{\tau}

\def\s{\sigma}
\def\th{\theta}

\def\dd{\mbox{d}}

\def\O{\Omega}

\def\bra{\langle}
\def\ket{\rangle}
\def\a{\alpha}

\def\d{\delta}
\def\D{\Delta}

\def\e{\epsilon}

\def\f{\phi}

\def\vf{\varphi}

\def\l{\lambda}
\def\L{\Lambda}
\def\m{\mu}
\def\n{\nu}
\def\s{\sigma}

\def\r{\rho}
\def\t{\tau}
\def\th{\theta}

\newcommand{\tn}[1]{\mbox{\tiny #1}}
\renewcommand{\@}[1]{\sqrt{#1}}
\renewcommand{\le}[1]{\label{#1}\end{eqnarray}}
\newcommand{\bea}{\begin{eqnarray}}
\newcommand{\eea}{\end{eqnarray}}

\newcommand{\eq}[1]{(\ref{#1})}
\def\nn{\nonumber\\}

\def\ffract#1#2{\raise .35 em\hbox{$\scriptstyle#1$}\kern-.25em/
\kern-.2em\lower .22 em \hbox{$\scriptstyle#2$}}
\def\GN{G_{\mbox{\tn N}}}

\newcommand\mathC{\mkern1mu\raise2.2pt\hbox{$\scriptscriptstyle|$}
        {\mkern-7mu\rm C}}              
\newcommand{\mathR}{{\rm I\! R}}         

\begin{document}
\pagestyle{empty}

\rightline{MCTP-15-22}
\vskip 1 truecm

\centerline{{\Large \bf Conceptual Aspects of Gauge/Gravity Duality}}
\vskip 1truecm

\begin{center}
{\large Sebastian De Haro$^{1,2}$, Daniel R. Mayerson$^{3,4}$, Jeremy N.~Butterfield$^5$}\\
\vskip .7truecm
{\it $^1$Department of History and Philosophy of Science, University of Cambridge\\
Free School Lane, Cambridge CB2 3RH, United Kingdom}\\
{\it $^2$Amsterdam University College, University of Amsterdam,
Science Park 113\\ 1090 GD Amsterdam, The Netherlands}\\
{\it $^3$ Institute for Theoretical Physics, University of Amsterdam, Science Park 904\\
Postbus 94485, 1090 GL Amsterdam, The Netherlands}\\
{\it $^4$ Department of Physics and Michigan Center for Theoretical Physics,\\
University of Michigan, 450 Church Street, Ann Arbor, MI 48109-1020, USA}\\
{\it $^5$ Trinity College, Cambridge, CB2 1TQ, United Kingdom}

\vskip .5truecm
{\tt sd696@cam.ac.uk, drmayer@umich.edu, jb56@cam.ac.uk}
\vskip.3truecm
5 July, 2016

\vskip .5truecm 
\end{center}

\vskip 3.5truecm

\begin{center}

\textbf{\large \bf Abstract}
\end{center}
We give an introductory review of gauge/gravity duality, and associated ideas of holography, emphasising the conceptual aspects. The opening Sections gather the ingredients, viz.~anti-de Sitter spacetime, conformal field theory and string theory, that we need for presenting, in Section \ref{adscftintroduced}, the central and original example: Maldacena's AdS/CFT correspondence. Sections \ref{dict} and \ref{hologeneral} develop the ideas of this example, also in applications to condensed matter systems, QCD, and hydrodynamics. Sections \ref{dS?} and \ref{BHhology} discuss the possible extensions of holographic ideas to de Sitter spacetime and to black holes. Section \ref{philaspects} discusses the bearing of gauge/gravity duality on two philosophical topics: the equivalence of physical theories, and the idea that spacetime, or some features of it, are emergent.

\newpage
\pagestyle{plain}

\tableofcontents

\newpage

\section{Introduction}\label{intro}
In the last twenty years, gauge/gravity duality, and associated holographic ideas, have come to dominate fundamental physics. The basic idea is that a theory, typically a gravity theory,  defined on a $D$-dimensional space or spacetime (`the bulk') can  be equivalent to a theory, typically a gauge theory, defined on a $(D-1)$--dimensional space or spacetime that forms the bulk's boundary. (That the physics of the bulk is adequately encoded in the boundary is reminiscent of holography: hence the use of the term.)  This equivalence is expressed by a correspondence (`dictionary') between the concepts (in particular, states and quantities) of one theory, and those of the other. Much of the interest in such examples arises from the facts that:\\
\indent (i) the concepts that the dictionary declares to `translate' into each other can be strikingly different: for example, a direction in the bulk spacetime is translated as the direction of the renormalization group flow in the boundary theory;\\
\indent (ii)  one theory's strong coupling regime, where problems are hard to solve by perturbative methods, can be translated into the other theory's weak coupling regime, where problems are easier to solve; and {\em vice versa}: suggesting that each theory can be used to help solve problems in the other.   

The aim of this paper is to give an introductory review of these developments, emphasising the conceptual aspects.

 The original example, conjectured by Maldacena (1998), concerns a gravity theory (a string theory) on an anti-de Sitter spacetime, being equivalent to a gauge theory (a conformal field theory) on its boundary: hence the name, `AdS/CFT conjecture'. More precisely, Maldacena conjectured that type IIB string theory on a AdS$_5\times S^5$ background is equivalent to  $\mathcal{N}=4$ four-dimensional super-Yang Mills field theory. This example has been much studied and has been very fruitful; the opening Sections of the paper gather the  ingredients that we need in order to present it. We stress that although there is by now a great deal of evidence in favour of it, still it remains a conjecture.\footnote{That it is hard to prove is of course hardly surprising, since it links weak and strong coupling regimes. Green~(1999), section 8, deals with a number of quantum corrections. For a discussion of some non-perturbative results in gauge/gravity dualities, see De Haro et al.~(2016: Section 4). Recent progress in non-perturbative calculations is reported in e.g.~Codesido et al.~(2015), and references therein.} We will present anti-de Sitter spacetime (Section \ref{genpro}), conformal field theory (Section \ref{qcftrecap}) and string theory (Section \ref{stringrecap}); and then in Section \ref{adscftintroduced}, the  AdS/CFT conjecture itself. 
 
 Sections \ref{dict} and \ref{hologeneral} develop the ideas of this example, also in application to condensed matter systems, QCD, and hydrodynamics. Sections \ref{dS?} and \ref{BHhology} discuss the possible extensions of holographic ideas to de Sitter spacetime and to black holes. We stress that, again in these Sections, the dualities discussed are (for the most part)  heuristic and-or conjectural.  Section \ref{philaspects} discusses the bearing of gauge/gravity duality on two philosophical topics: the equivalence of physical theories, and the idea that spacetime, or some features of it, are emergent. In this Section (and implicitly throughout the paper), we will take a theory to be given by a state-space, equipped with various structures, especially a set of quantities and a dynamics; and a duality to be a bijective structure-preserving mapping between theories thus understood. Two theories being equivalent will then be a matter of their `saying the same thing' under the duality mapping. Section \ref{concl} concludes. 

Thus, broadly speaking, the first half of the paper develops the AdS/CFT conjecture in enough detail to articulate the conceptual issues involved. The second half of the paper considers applications and generalizations of the ideas developed in the first half: especially to spacetimes other than anti-de Sitter,  and to black holes. 

We should stress two main limitations of our discussion. (i): We downplay the many `spin-offs' in condensed matter physics and other fields that gauge/gravity duality has spawned: in effect, we restrict ourselves to cases closely related to AdS/CFT---to classical conformal symmetry and cases close to it.  (ii): Obviously, there is a large literature on the topic of each Section, and  research on each topic is still ongoing. We do not in any way aim to give a complete survey of the field, which would require a book. Happily, there now {\em is} an excellent book: Ammon and Erdmenger (2015). Let us also here mention some earlier excellent pedagogic expositions: by Maldacena himself (2003, 2004, 2004a); and by others, Horowitz and Polchinski (2006), McGreevy (2010), Petersen (1999), Skenderis (2002).

The paper can be read in two ways. (i): Readers wishing to learn some of the physics involved,  as well as the conceptual issues, can read the paper linearly. (ii): Readers interested mainly in the conceptual and philosophical issues that arise in gauge/gravity dualities (or dualities in general---since gauge/gravity dualities exemplify many general issues) can read the introductory sections and then skim through later Sections. Sections 6.2., 10 and 11 contain the philosophical meat.

\section{AdS spacetime}\label{genpro}

Anti-de Sitter spacetime is a maximally symmetric spacetime with negative curvature and Lorentzian signature. A maximally symmetric spacetime of dimension $D$ is one that has the maximum number of Killing vectors, $D(D+1)/2$: which also implies that the spacetime has constant curvature. For a maximally symmetric spacetime, the Riemann tensor can be written as:\footnote{For a more detailed exposition of the geometrical properties of AdS, see Ammon and Erdmenger (2015: Section 2.3).}
\bea\label{Rmn}
R_{\m\n\l\s}=-{1\over\ell^2}\left(g_{\m\l}g_{\n\s}-g_{\m\s}g_{\n\l}\right),
\eea
where $\ell$ is called the `AdS radius of curvature'. One can show that for a spacetime satisfying \eq{Rmn} the Weyl tensor identically vanishes and the space is conformally flat. Contracting indices, one also readily sees that the solutions of \eq{Rmn} are also solutions of Einstein's equations in $D$ dimensions with cosmological constant $\L=-{(D-1)(D-2)\over2\ell^2}$.

A $D$-dimensional space of constant curvature is best constructed as a slice of an embedding  $(D+1)$-dimensional flat space. Think, for instance, of how a two-sphere can be seen as a submanifold of $\mathbb{R}^3$. In a similar way, we can construct $D$-dimensional anti-de Sitter spacetime as a hyperboloid in a $(D+1)$-dimensional embedding Minkowski space with two time directions:
\bea\label{embed}
-X_0^2-X_D^2+\sum_{i=1}^{D-1}X_i^2=-\ell^2~.
\eea
Solving  this constraint, and calculating the induced metric on the surface from the flat $(D+1)$-dimensional Minkowski metric, we get the AdS metric. It is not hard to find a system of coordinates $(\r,\t,\O_i)$ that covers the entire hyperboloid:
\bea\label{sol}
X_0&=&\ell\cosh\r\,\cos\t\nn
X_D&=&\ell\cosh\r\,\sin\t\nn
X_i&=&\ell\sinh\r~~\O_i~~(i=1,\ldots,D-1)~,
\eea
where $\Omega_i$ is a unit vector parametrising a $(D-2)$-sphere, satisfying $\sum_{i=1}^{D-1}\Omega_i^2=1$. This is analogous to solving the constraint for the two-sphere in ${\mathbb R}^3$, where the sines and cosines in one of the angles are replaced by the corresponding hyperbolic quantities in \eq{sol}, owing to the non-compactness of the hyperboloid \eq{embed}. The range of the coordinates is thus $\r\in[0,\infty)$, $\t\in[0,2\pi)$. The AdS line element 
\bea\label{global}
\dd s^2=\ell^2\left(-\cosh^2\r\,\dd\t^2+\dd\r^2+\sinh^2\r\,\dd\O_{D-2}^2\right).
\eea
is obtained by substituting the above constraint in the embedding flat Minkowski metric.

This way of constructing AdS has the advantage that the symmetry group is apparent from the defining equation \eq{embed}: namely, $\mbox{SO}(2,D-1)$, the symmetry group of the hyperboloid. From the solution \eq{sol}, one also readily sees how this symmetry group leaves the metric invariant.

Euclidean AdS space, which is most often used in work on  AdS/CFT, can be constructed in the same way. We simply change one of the minus signs in \eq{embed} and modify \eq{sol} accordingly. The resulting line element then has Euclidean signature. 

The following transformation makes the metric conformal to $\mathbb{R}\times S^{D-1}$: $\tan\th:=\sinh\r$, so that
\bea\label{conf}
\dd s^2={\ell^2\over\cos^2\th}\left(-\dd\t^2+\dd\th^2+\sin^2\th\,\dd\O_{D-2}^2\right).
\eea
Notice that $\th$ only has half of the range: $[0,{\pi\over2})$ instead of $[0,\pi)$. Therefore, we do not quite have a full $S^{D-1}$ here, but only its northern hemisphere. As one approaches the equator, $\th\rightarrow{\pi\over2}$, the line element \eq{conf} diverges; this is no problem because the spacetime is non-compact and the equator was not part of the spacetime to begin with (notice the semi-open interval for $\th$). Whereas the rest of the line element is completely regular, the conformal factor (roughly representing the volume of the space) diverges as one approaches the boundary  $\th\rightarrow\pi/2$. The boundary is timelike and corresponds to the region $\r\rightarrow\infty$ in which all of the $X$'s in \eq{sol} diverge. This is precisely the asymptotic infinity of the hyperboloid, as expected. The conformal map thus maps the asymptotic infinity to a finite parameter distance. 

In Minkowski space, which is conformally mapped to a diamond with lightlike boundaries, it is enough to specify the fields on a spatial slice. This slice is causally connected, by means of null rays, to the entire spacetime, and the values of the fields on the slice determine the time evolution. As we have seen, however, the AdS boundary is not lightlike, but timelike: in order to have a well-defined Cauchy problem, we need to specify the boundary conditions at spacelike infinity rather than on some spatial surface---since light rays reach the boundary in finite time. As we shall see in Section \ref{dict}, the specification of the boundary conditions is one of the key ingredients in the AdS/CFT dictionary.

At spacelike infinity, the line element \eq{conf} is conformal to the standard line element on $\mathbb{R}\times S^{D-2}$. This will turn out to be the metric of the CFT, and the conformal factor is indeed irrelevant because of the conformal symmetry of the CFT. 

Since the metric in the CFT is the induced boundary metric up to a conformal factor, it is very convenient to write the metric in a form that is conformally flat. This is achieved with the following coordinate transformation:
\bea\label{poincare}
X_0&=&{1\over2r}\left(r^2+\ell^2+g_{ij}\,x^ix^j\right)\nn
X_D&=&{1\over2r}\left(r^2-\ell^2+g_{ij}\,x^ix^j\right)\nn
X_i&=&{\ell\over r}\,x_i~,~~i=1,\ldots,D-1~,
\eea
where $g_{ij}$ is the $(D-1)$-dimensional Lorentzian or Euclidean flat metric: $g_{ij}\, x^i x^j=-t^2+\sum_{i=1}^{D-2} x_i^2$ in the Lorentzian case, and $g_{ij}\,x^ix^j=\sum_{i=1}^{D-1}x_i^2$  in the Euclidean case. Since $r$ will play a key role in what follows (as: (i) parametrising the boundary; (ii) governing RG flow) we invert the expression for it in \eq{poincare}:
\bea\label{invertr}
r={\ell^2\over X_0-X_D}~.
\eea
One easily checks that the Lorentzian case is a solution of \eq{embed}, and the Euclidean case a solution of the same equation with only one negative sign on the left-hand side. These coordinates only cover a local patch of the spacetime, with $r>0$. As we noticed after \eq{conf}, the boundary is the asymptotic region of the hyperboloid \eq{embed}. We see from either \eq{poincare} or \eq{invertr} that this corresponds to $r\rightarrow0$. 
 The metric is then:
\bea\label{metric}
\dd s^2={\ell^2\over r^2}\left(\dd r^2+g_{ij}\,\dd x^i\dd x^j\right)~,
\eea
The vector $x^i$ $(i=1,\ldots,d:=D-1)$ parametrises the boundary coordinates since the boundary is now at $r\rightarrow0$. As we see, this metric is conformal to the flat metric on $\mathbb{R}^D$ (again, only half of it because $r>0$), with either Lorentzian or Euclidean signature. The boundary metric is then the flat metric on $\mathbb{R}^{D-1}$. These coordinates are called {\em Poincar\'e coordinates}, as they represent a higher-dimensional version of the Poincar\'e half-plane (cf.~Farkas et al.~(1980), pp.~213ff.)

\section{Quantum and Conformal Field Theories}\label{qcftrecap}

Quantum field theories\footnote{Some lecture notes on conformal field theories are Gaberdiel (1999), Ginsparg (1988), or Cardy (2001); see e.g.~Ketov (1995) for lecture notes on \emph{super}conformal field theories.} are theories that describe the dynamics of \emph{fields}, which are objects that permeate all of spacetime. The elementary quanta of excitation of these fields can then be interpreted as particles. For example, electrons can be thought of as the ripples of the electron field.

Relativistic quantum field theories  enjoy, by definition, Poincar\'e invariance: that is, invariance under the  transformations
\begin{align} 
\label{eq:translation} x^{\mu} &\rightarrow x^{\mu} + a^{\mu}~,\\
\label{eq:rotation} x^{\mu} &\rightarrow M^{\mu}_{\ \nu}\, x^{\nu}~,
\end{align}
representing translations with constant parameter $a^{\mu}$ and rotations/Lorentz boosts with Lorentz transformation matrix $M$. The operators that generate the infinitesimal\footnote{The infinitesimal transformation on the coordinates is given by:
$ x'^{\mu} = x^{\mu} + \omega_a\, \frac{\partial x^{\mu}}{\partial \omega_a}\,,$
where $\omega_a$ are infinitesimal parameters. The operator $O_a$ generating the transformation of the field $\phi$ to $\phi'$ is defined as the operator that acts on $\phi$ such that:
$ \phi'(x) = \phi(x) - i\omega_a \left(O_a \phi(x)\right).$} translations and rotations (and boosts) are $P_{\mu}$ and $L_{\mu\nu}$; for example,\footnote{In units where $\hbar=c=1$, which we use throughout this paper.} $P_{\mu} = -i\,\partial_{\mu}$.

 Conformal field theories are, in short, theories that are invariant under a larger group of spacetime transformations called \emph{conformal transformations}. The group of conformal transformations is generated by the Poincar\'e transformations (\ref{eq:translation}) and (\ref{eq:rotation}), as well as the two additional transformations:
\begin{align}
 \label{eq:dilatation} x^{\mu} &\rightarrow \lambda\, x^{\mu}~,\\
 \label{eq:specialCT} x^{\mu} &\rightarrow \frac{x^{\mu}-b^{\mu} x^2}{1-2b\cdot x + b^2 x^2}~,
\end{align}
respectively called \emph{dilatations} (with parameter $\lambda$) and \emph{special conformal transformations} (with parameters $b^{\mu}$). Dilatations are easy to visualize as simply `stretching out' spacetime; special conformal transformations are less easy to visualize but can be thought of as resulting from the effect of `inverting' the spacetime coordinates ($x^{\mu}\rightarrow x^{\mu}/x^2$) twice, with a translation in between. An alternative but equivalent formulation of conformal invariance is invariance under all coordinate transformations $x\rightarrow x'(x)$ such that the metric transforms as:
\be g_{\mu\nu}(x) \rightarrow g_{\mu\nu}'(x') = \Lambda(x)\, g_{\mu\nu}(x),\ee
for an arbitrary function $\Lambda(x)$; Poincar\'e invariance would be the restriction to $\Lambda=1$.

The operators or fields in a relativistic quantum field theory\footnote{Typically, operators/fields in a relativistic quantum field theory are referred to as \emph{fields}; confusingly, the convention is to use the name \emph{operator} when talking about operators/fields in conformal field theories.} must fall into representations of the Poincar\'e group. This means we can choose a basis of fields in the theory that are eigenfunctions of the Lorentz operators $L_{\mu\nu}$.
For example, for such a field $\phi$ at the origin:
\be
 [L_{\mu\nu}, \phi(0)] = S_{\mu\nu}\, \phi(0),
\ee
where we call $S$ the \emph{spin matrix} of the field $\phi$. Similarly, in a conformal field theory, operators fall into representations of the conformal group; we can choose a basis of operators that are eigenfunctions, not only of the Lorentz operators $L_{\mu\nu}$, but also of the dilatation operator $D$ and the special conformal transformation operator $K_{\mu}$. Again, using an operator $\phi$ as an example:
\begin{align}
 \label{eq:L} [L_{\mu\nu}, \phi(0)] &= S_{\mu\nu}\, \phi(0),\\
  \label{eq:D} [D,\phi(0)] &= -i\Delta\, \phi(0),\\
 \label{eq:K} [K_{\mu},\phi(0)] &= \kappa_{\mu}\, \phi(0).
\end{align}
The number $\Delta$ is called the \emph{scaling dimension} (or simply dimension) of $\phi$. 

The algebra of conformal transformations implies that all operators are either \emph{primary} with $\kappa_{\mu}=0$ or \emph{descendants} if $\kappa_{\mu}\neq0$.
All descendants can be thought of as `derivatives' of primary operators, in the sense that they can be written (schematically) as $\partial^n \phi$ for some primary operator $\phi$. Using conformal invariance, the properties of descendant operators follow more or less immediately from those of the primaries; this means that a study of conformal field theories is typically a study of its primary operators. Gauge invariant primary operators in a CFT are completely classified by their quantum numbers $S_{\mu\nu}$ and $\Delta$.

One can imagine field theories that are invariant under an even larger group of transformations than conformal symmetry: namely, by including supersymmetry. These \emph{superconformal} field theories are invariant under the (bosonic) conformal transformations $P_{\mu},L_{\mu\nu},D,K_{\mu}$ as well as the (fermionic) supersymmetry transformations $Q$ and superconformal transformations $S$. These fermionic operators $Q,S$ will carry a spinor index $\alpha$ as well as an internal index $A$. The latter runs from 1 to $\mathcal{N}$; for $\mathcal{N}=1$ we have a theory with \emph{minimal} supersymmetry, while for $\mathcal{N}>1$ we have \emph{extended} supersymmetry. When $\mathcal{N}>1$ there are always additional (bosonic)  symmetries, called \emph{$R$-symmetries}. The $R$-symmetry generators commute with the (bosonic) conformal symmetry generators and have the effect of rotating the supercharges' internal index $A$. 
In theories with minimal supersymmetry, we can apply the supersymmetry transformation to each field once, and obtain its supersymmetric partner. When there is extended supersymmetry, starting with a particular field we can apply ${\cal N}$ supersymmetry transformations to obtain each time a new field. Theories with extended supersymmetry thus contain more fields and higher spin values, because the supercharge increases the spin by 1/2.

 Again, operators in a superconformal field theory fall into representations of the superconformal algebra, which are composed of multiple representations of the conformal algebra. A \emph{superconformal primary} operator is an operator $\phi$ which satisfies:
\begin{align}
 [K_{\mu},\phi(0)] &=0,\\
 [S,\phi(0)] &=0.
\end{align}
Certain special superconformal primary operators called \emph{chiral primary} operators have special properties, such as having their scaling dimension $\Delta$ protected from quantum corrections---this means one can compute $\Delta$ at small coupling (e.g.~using perturbation theory) and be assured the result is valid at large coupling as well.

In a relativistic quantum field theory, one is typically interested in studying the $S$-matrix, which describes the scattering of particles. However, to be able to meaningfully define an $S$-matrix, one needs to be able to `separate' the particles at temporal and spatial infinity, so as to give the initial and final states of the scattering process. In a conformal field theory, we can never really separate particles in such a way at infinity because of the conformal invariance---intuitively, this is because we can always use a dilatation transformation to bring the particles closer together again. Hence, for conformal field theories, there is no meaningful way to define a $S$-matrix. Instead, one is forced to study the correlation functions of operators $\langle \mathcal{O}_1\cdots \mathcal{O}_n\rangle$. While perhaps less intuitively satisfying than the $S$-matrix description, correlation functions contain all of the physical information of a field theory (just like the $S$-matrix does).


Conformal invariance greatly restricts the possible form of correlation functions of operators. The two-point function of two primary operators, $\mathcal{O}_1$ and $\mathcal{O}_2$, is forced to be:
\be 
\langle \mathcal{O}_1(x)\, \mathcal{O}_2(y)\rangle = \begin{cases} |x-y|^{-2\Delta_1}, & \text{if $\Delta_1=\Delta_2$},\\
                                                         0 &  \text{otherwise;}
                                                       \end{cases}
\ee
so it is completely determined by the scaling dimensions $\Delta_{1,2}$ of the scalar operators $\mathcal{O}_{1,2}$. Similarly, the three-point function of primaries is restricted to be:
\be \langle \mathcal{O}_1(x) \mathcal{O}_2(y) \mathcal{O}_3(z)\rangle = C_{123}\,\times~|x-y|^{-\Delta_{12}}\, |y-z|^{-\Delta_{23}}\, |x-z|^{-\Delta_{13}}~,\ee
where $\Delta_{ij} = \Delta_i + \Delta_j - \Delta_k$ (with $k$ defined as $\neq i$ and $\neq j$). We see that the three-point function depends on a real-number coefficient $C_{123}$ as well as the scaling dimensions of the operators; this coefficient $C_{123}$ contains dynamical information about the theory. It can be proven that all of the four- and higher-point functions in a CFT are essentially fixed by conformal invariance once the scaling dimensions and three-point coefficients $C_{ijk}$ are known. Thus, the physics of operators in a CFT is completely fixed by specifying its (primary) operator content, their scaling dimensions, and all of their three-point coefficients.

Another important feature of conformal field theories is the \emph{state-operator correspondence}. This is the fact that any state in a CFT is in one-to-one correspondence with an operator of the CFT that is local, i.e. defined at each point. Loosely speaking, this correspondence tells us that every state in the theory can be obtained by acting with a local operator on the vacuum. For example, the vacuum itself trivially corresponds to the identity operator acting on the vacuum. The state-operator correspondence implies that complete knowledge of the spectrum of primary operators in a CFT is equivalent to knowing the full spectrum of states in the theory.

One superconformal field theory that is of central importance when discussing AdS/CFT is four-dimensional $\mathcal{N}=4$ $\mbox{SU}(N)$ super Yang Mills (SYM) theory. $\mbox{SU}(N)$ Yang-Mills is the field theory of a gauge field with non-Abelian gauge group $\mbox{SU}(N)$; one can then couple 6 (real) scalars and 4 (complex Weyl) fermions to the gauge field to obtain a theory with $\mathcal{N}=4$ extended supersymmetry. The resulting complicated, supersymmetric gauge theory can be proven to be exactly conformal. All of the relative interactions between the gauge field, scalars, and fermions are completely fixed by supersymmetry; the only overall tunable parameter in the theory is the gauge coupling $g_{\tn{YM}}$ (and a theta angle $\theta_{\tn{YM}}$).

Studying Yang-Mills theories (non-Abelian gauge theories) is, in general, very hard. The prime example is quantum chromodynamics (QCD), the theory of strong interactions, which is $\mbox{SU}(3)$ Yang-Mills theory coupled to fermions. QCD is confining at low energies---we only see baryons (e.g.~protons) and mesons (e.g.~pions) instead of loose quarks and gluons. Confinement is an extremely difficult phenomenon to study, since it is an inherently strong-coupling phenomenon and thus not accessible to the usual perturbation techniques used in field theory. $\mathcal{N}=4$ SYM is not confining (since it is conformal and thus does not have an energy-dependent physics) but is nevertheless difficult to study if the gauge coupling $g_{\tn{YM}}$ is large. However, 't Hooft (1974) realized Yang-Mills theories simplify when the gauge group rank $N$ is taken to be very large. In the \emph{'t Hooft limit}, one formally takes $N\rightarrow \infty$ while keeping the \emph{'t Hooft coupling} $\lambda := g_{\tn{YM}}^2 N$ fixed. In this limit, it can be shown that all Feynman diagrams except so-called \emph{planar diagrams} contribute (to, for example, correlation functions) at only {\em sub}-leading order in $N$ and thus can be ignored---greatly simplifying the theory.\footnote{For an introduction of the 't Hooft limit, see section 1.7.5 of Ammon et al.~(2015). Bouatta and Butterfield (2015) is a conceptual discussion of the 't Hooft limit.}

\section{String Theory}\label{stringrecap}


While we aim to introduce the AdS/CFT correspondence with as little string theory background as possible, it is useful to briefly review a few core concepts of string theory that are essential to understand the conjecture and its motivation.\footnote{There are numerous books that introduce string theory at various levels. Canonical and fairly comprehensive volumes introducing string theory at the graduate level include Green, Schwarz, Witten (1987) and Polchinski (1998). A volume also suitable for advanced undergraduates due to its more pedagogical approach is e.g.~Zwiebach (2009). There are also various excellent lecture notes on string theory, e.g.~Tong (2009).}

String theory, as its name suggests, is basically a theory of strings: one-dimensional extended (rather than pointlike) objects; when a string moves, it sweeps out a {\it world-sheet} rather than a {\it world-line}. Strings can be closed (i.e. a loop) or open; the closed string massless excitations can be shown to include e.g.~gravity, while open string excitations include e.g.~gauge field dynamics. The strings can only interact by combining or splitting, with an interaction strength set by the closed \emph{string coupling} $g_s$. There is also an open string coupling $g_{\tn{o}}$, but this is simply related to the closed string coupling as $g_s\sim g_{\tn{o}}^2$---essentially this is because two open strings can combine into (or arise from the splitting of) one closed string. It is interesting to note that $g_s$ is not a parameter of string theory that needs to be fixed externally;\footnote{Indeed, the only fundamental tunable parameter in string theory is $\a'=l_s^2$, the string length squared.
} 
$g_s$ is set dynamically by the expectation value of the dilaton $\phi$, a massless scalar field in the closed string spectrum, through $g_s = \langle e^{\phi}\rangle$.

Besides strings, string theory also contains other objects called Dirichlet-branes\footnote{The name refers to the fact that open strings that end on D-branes have \emph{Dirichlet} boundary conditions, which mean that the string endpoints are stuck on the branes.} or \emph{D-branes}. 
A D-brane of a specific dimension is also called a D$p$-brane, where $p$ indicates that the brane spans $p$ spatial dimensions (and thus has a $(p+1)$ dimensional spacetime {\it world-volume}).

D-branes are essentially objects on which open strings have their endpoints. The dynamics of the open string endpoints is then equivalent to the dynamics of the D-branes themselves, making the branes truly dynamical objects in their own right. The excitations of the open string endpoints transverse to the brane translate into fluctuations of the brane itself in the ambient spacetime, while excitations of the open strings parallel to the brane give dynamics for a gauge field
  living in the world-volume of the brane itself---this gauge field becomes non-Abelian when multiple D-branes are stacked on top of each other. The gauge field coupling constant is related to the string coupling constant as $g_{\tn{YM}}^2\sim g_s$; (again, this is because $g_s$ is the \emph{closed} string coupling and we need two open strings with coupling constant $g_{\tn{o}}\sim g_{\tn{YM}}$ to be able to form one closed string). Which gauge field describes the dynamics of a particular D-brane depends on the type of D-brane considered. For example, for $N$ coincident D3-branes, the relevant world-volume gauge theory is four-dimensional $\mathcal{N}=4$ $\mbox{SU}(N)$ super Yang Mills theory.\footnote{Strictly speaking, the gauge group is $\mbox{U}(N)$ instead of $\mbox{SU}(N)$. The $\mbox{U}(N)$ gauge theory is essentially equivalent (up to global $\mathbb{Z}_N$ identifications) to a free $\mbox{U}(1)$ gauge field and a $\mbox{SU}(N)$ gauge theory. The $\mbox{U}(1)$ part is related to the motion of the center of mass of the stack of D3-branes and can typically be considered as decoupled from the $\mbox{SU}(N)$ gauge theory.}

Two of the most studied versions of string theory are the maximally supersymmetric type IIA and type IIB string theories in 10 spacetime dimensions. Both theories contain closed strings as well as D-branes of different dimension (and thus also open strings). IIA contains D$p$-branes with $p$ even while IIB contains D$p$-branes with $p$ odd.

Type IIA and IIB string theories are very closely related by {\em $T$-duality}. $T$-duality is an inherently \emph{stringy} phenomenon that occurs when a spatial direction is compact. It relates type IIA (resp. IIB) string theory compactified on a spatial direction with radius $R$ to type IIB (resp. IIA) string theory compactified on a spatial direction with radius $\alpha'/R$ ($\alpha'=l_s^2$ is the square of the fundamental string length). The winding modes of IIA (resp. IIB) strings along this compact direction are mapped to momentum modes of the IIB (resp. IIA) string along the transformed direction. Finally, D$p$-branes of one theory are mapped to D$(p\pm1)$-branes of the dual theory.

Studying the full dynamics of string theory is a very hard problem, so often a low-energy limit is used to simplify things. In such a low-energy limit, one can essentially neglect the effects of the strings' finite length and approximate them by point particles. The low-energy limit of the closed string sector of type IIA/B string theory is described by type IIA/B supergravity, a supersymmetric theory of Einstein gravity in 10 dimensions coupled to a number of extra fields. In this low-energy supergravity description, we can construct and study solutions (such as black holes) of the theory just as we would in familiar classical gravity theories. For example, this low-energy limit also explicitly contains strings\footnote{Note that the strings we can describe in solutions of supergravity are a bit different from the closed strings that make up the background (about which we have just said we can ignore their finite length!); heuristically, these strings described by supergravity solutions are \emph{longer} strings, extending over macroscopic lengths.} and D-branes, which are represented by particular solutions of the supergravity theory. Such supergravity solutions can only be trusted as long as they stay within the low-energy limit. This is the case as long as $\alpha'/\ell^2\ll 1$, where $\alpha'=l_s^2$ is the string length (squared) and $\ell$ is the (local) radius of curvature of the supergravity solution---if this constraint is violated, the finite string length effects cannot be neglected consistently and the low-energy supergravity approximation breaks down.

\section{The AdS/CFT Duality Introduced}\label{adscftintroduced}


To introduce the AdS/CFT correspondence, we will in this Section present its first and most-studied example. We will do this in a mostly heuristic fashion, following the ideas in the original presentation by Maldacena (1998). Aharony et al.~(2000) is a comprehensive early presentation.

 Namely, this example is the duality between:\\
 \indent  \indent (i) $\mathcal{N}=4$ four-dimensional (4D) super Yang Mills (SYM) field theory (introduced in Section \ref{qcftrecap}); 
 and \\
 \indent \indent (ii) type IIB string theory on an AdS$_5\times S^5$ background (cf.~Sections 2 and 4).

This duality arises from describing the 
low energy dynamics of a stack of $N$ coincident D3-branes in a flat background in type IIB string theory---in two different ways. Here $N$ is, for the moment, just an integer: the number of coincident D3-branes. We will shortly see that, according to the AdS/CFT duality, it corresponds to the rank $N$ of the gauge group $\mbox{SU}(N)$; and that it must be taken to be large, as in the 't Hooft limit at the end of Section \ref{qcftrecap}. 

First of all, let us consider the low-energy limit, where we only keep the massless states, and the dynamics of the D-branes decouples from the ambient spacetime: the excitations far away from the brane completely decouple from the excitations close to the brane---this is called the \emph{decoupling limit}.

As we said in Section \ref{stringrecap}, the dynamics of the D-branes in this low-energy limit will be entirely described by the excitations of the open string endpoints on the D-branes. This gives a field theory that lives on the four-dimensional (4D) world-volume of the D3-branes: in this case, it will be $\mathcal{N}=4$ 4D super-Yang-Mills theory  (SYM), a much-studied  superconformal field theory.

On the other hand, it is known that there exists a supergravity solution that describes D3-branes (see the discussion at the end of Section \ref{stringrecap}).
When we zoom in to the region close to the D3-branes, this supergravity solution precisely becomes AdS$_5\times S^5$. This procedure of `zooming in' is also a \emph{decoupling limit} and represents the fact that 
the excitations near the branes (i.e. those in the AdS space) decouple completely from the excitations in the (asymptotically flat) region far away from the branes.
Note that in this AdS$_5$ description, the D3-branes are considered to be `dissolved'
into the closed string sector. This means that the branes' effects remain and are taken into account in the curved geometry; but the branes themselves and the relevant open string degrees of freedom (e.g.~the gauge field on the brane) can no longer be seen explicitly in the (closed string) geometry. This is sometimes called a \emph{geometric transition} of the branes.

The AdS/CFT duality (also known as: correspondence) is now simply the conjecture that---even when the coupling is {\em not} low---these two theories are equivalent: (i) $\mathcal{N}=4$ SYM and (ii) string theory on AdS$_5\times S^5$. 

But if the two theories are indeed equivalent, calculations in one theory should in principle give the same result as calculations in the other. It is this that gives the conjectured equivalence such promise. It relates the regime where one theory is hard to calculate in, to the regime where the other theory is easy (well: easier!) to calculate in; as follows. Calculations in the AdS$_5\times S^5$ supergravity solution can only be trusted as a good approximation to the full string theory when the characteristic length scale of the solution is large compared to the string scale. But the radii of both the $S^5$ and the AdS$_5$ factor spacetimes are proportional to $g_s N$; where, as before, $g_s$ is the string coupling and $N$ is the number of coincident D3-branes in the original D3-brane picture (i.e.~before the decoupling limit is taken). So this translates into the condition that $g_s N \gg 1$. That is: if this condition is not satisfied, we cannot trust the supergravity approximation; we must take stringy corrections into account. On the other hand, the relevant parameter in $\mathcal{N}=4$ super-Yang-Mills theory, with which we can do perturbation theory, is also proportional to $g_s N$. Remember from section 4 that $g_s\sim g_{\tn{YM}}^2$, so this is the 't Hooft parameter. This means that we can only trust perturbation theory calculations when $g_s N \ll 1$, i.e.~of weak 't Hooft coupling; otherwise non-perturbative effects become important.

Here, as an important aside, we should also mention another restriction on the number $N$. It is important for the above discussion that $N$ is large---in fact, strictly speaking, the correspondence as described above is only valid in the limit $N\rightarrow \infty$ (cf.~the discussion of the 't Hooft limit in gauge theories at the end of Section \ref{qcftrecap}). In principle, one can calculate $1/N$ corrections to the correspondence described above: for example, these correspond to quantum corrections in the AdS$_5$ geometry, which only in the $N\rightarrow\infty$ limit is captured by classical supergravity. Thus, we emphasize the (somewhat subtle) difference in the bulk between \emph{stringy} corrections, which are \emph{classical} corrections due to the fact that a string has finite size (as opposed to pure gravity where the graviton is a point particle of zero size) and are suppressed if $g_s N\gg 1$; and \emph{quantum} corrections, which are due to the quantum nature of string theory and are suppressed if $N\gg 1$.


We see that the regime where we can trust the perturbation calculations in the field theory ($g_s N\ll 1$) and where we can trust the AdS$_5$ supergravity solution ($g_s N\gg 1$) do not overlap. This is {\em good news}. For by studying a supergravity solution, we can gain insight into the very difficult non-perturbative physics of strongly coupled $\mathcal{N}=4$ SYM. Conversely, perturbative calculations in this theory can---once `translated' to the string theory on the AdS$_5\times S^5$ background---give us results going beyond the supergravity approximation. 

This complementarity of regimes where we can reliably perform calculations makes the AdS/CFT correspondence an extremely powerful tool. But it also means it is very difficult to actually \emph{prove} the correspondence, by performing calculations in both descriptions at a fixed coupling and comparing them.

To close this heuristic introduction of AdS/CFT, let us mention a few interesting entries in the AdS/CFT correspondence's `dictionary', i.e. its mapping between the theories: specifically for the above example of AdS$_5\times S^5$ and $\mathcal{N}=4$ SYM. First of all, one can see that the spacetime symmetries of the gravity solution correspond to symmetries in the field theory. The spacetime symmetry transformations of AdS$_5$ correspond to the conformal symmetry transformations of the dual field theory. Indeed, as mentioned in Section \ref{genpro}, the symmetry group of AdS$_5$ is $\mbox{SO}(2,4)$ (which is the symmetry group of the quadric \eq{embed}). But $\mbox{SO}(2,4)$ is also the conformal group in four dimensions: in the language of Section \ref{qcftrecap}, this is the extension of the Poincar\'e group with the two additional generators $D$ and $K_{\mu}$; (cf.~\eq{eq:dilatation}-\eq{eq:specialCT} and \eq{eq:D}-\eq{eq:K}). On the other hand, the $\mbox{SO}(6)$ symmetry generators of $S^5$ correspond to the $R$-symmetry (see Section \ref{qcftrecap}) transformations of the field theory (see e.g.~Maldacena (1998), and Aharony et al.~(2000), Section 3). 

A second example is the correspondence of the radial coordinate $r$ in AdS$_5$ (introduced in \eq{metric}), which corresponds to a coarse-graining parameter of the renormalization group (RG) flow in the field theory, in the sense that moving along this coordinate further away from the boundary is like going to longer wavelengths and thus smaller energies in the field theory (see e.g.~de Boer, Verlinde, Verlinde (2000)). 

Finally, we note that the $n$-point correlation functions of the CFT (again, see the discussion in Section \ref{qcftrecap}) can be calculated by a very explicit procedure in the AdS spacetime by a method of connecting the points in the correlation function in the AdS spacetime by \emph{bulk-to-boundary propagators} (see e.g.~Witten (1998a), Section 2)). We will also touch on this point in Section \ref{dict} when we discuss the precise construction of the dictionary and how to calculate correlation functions holographically.

\section{Development}\label{dict}

In the previous Section we presented the AdS/CFT correspondence in the heuristic way that goes back to the work of Maldacena (1998). The correspondence as presented in Section \ref{adscftintroduced} is valid only when the metric is exactly AdS as in \eq{metric} or is very close to it, and the additional curvature of the geometry produced by the fields can be neglected. In this Section, we describe progress in the search for a more general and more rigorous formulation of the AdS/CFT correspondence. Such a formulation will have to specify, at the very least:\\
\indent 1) What are the physical quantities that characterize both sides of the duality? \\
\indent 2) How are these quantities related to each other by the duality? \\
These questions will be the focus of Section \ref{dictionary}. Then in Section \ref{backindpdce}, we will focus on whether a natural requirement for a theory of gravity, viz. background-independence, is satisfied.

\subsection{AdS/CFT in more detail}\label{dictionary}

We take a theory to be given by a state-space, equipped with various structures, especially a set of quantities\footnote{Our word `quantities' denotes what, in the physics literature, especially in quantum mechanics, are often called `observables'. We follow John Bell (1990) in regarding the latter as a `bad word'.} and a dynamics; and we take a duality to be a bijective structure-preserving mapping between  theories thus understood. This will be made more precise in Section \ref{equivtthies}. But we can already apply these construals to  AdS/CFT: beginning on the gravity or bulk side, and then considering the CFT/boundary theory. We will also first consider the vacuum case, i.e. pure gravity (Section \ref{bulk} and \ref{bulkbdy}); and then consider how the correspondence can incorporate matter fields (Section \ref{matter}).

\subsubsection{The bulk theory}\label{bulk}  
The states in the bulk are the states in a specific theory of quantum gravity: they consist of the configurations of the metric and matter fields that are compatible with the equations of motion. The quantities are the operators that are invariant under the symmetries. Usually, these are calculated perturbatively; e.g., by quantising the fluctuations about a pure AdS$_5$ solution. 

Let us first discuss the equations of motion and then get back to the operators. The low-energy approximation to the bulk equations of motion is given by Einstein's equations with a negative cosmological constant. As we will see shortly, the leading order in the low-energy approximation is enough to write down the most basic quantities.

So we need to solve Einstein's equations with a negative cosmological constant. This will give us the states. However, we are looking for general states, therefore the metric has to be much more general than \eq{metric}. 

It was shown by Fefferman and Graham (1985) that, for any space that satisfies Einstein's equations with a negative cosmological constant, and given a conformal metric at infinity, the line element can be written in the following form:
\bea\label{FG}
\dd s^2=G_{\m\n}\,\dd x^\m\dd x^\m={\ell^2\over r^2}\left(\dd r^2+g_{ij}(r,x)\,\dd x^i\dd x^j\right),
\eea
where $g_{ij}(r,x)$ is now an arbitrary function of the radial coordinate $r$. The remaining coordinates $x^i$ ($i=1,\ldots,d:=D-1$) parametrise the boundary, which is of dimension $d$ and, as in AdS, located at $r\rightarrow0$. The conformal metric at the boundary is $g_{(0)ij}(x):=g_{ij}(0,x)$. 

Solving Einstein's equations now amounts to finding $g_{ij}(r,x)$ given some initial data. As explained in Section \ref{genpro}, because of the presence of the timelike
boundary, choosing a spatial Cauchy surface at some initial time would not completely specify our problem. Instead, we have to provide boundary conditions. Thus it is best to specify the metric $g_{(0)ij}$ at the conformal boundary. Because Einstein's equations are second order, we also need to provide a second boundary condition for the metric. This is done as follows. Fefferman and Graham (1985) showed that $g_{ij}(r,x)$ has a regular expansion in a neighbourhood of $r=0$:
\bea\label{FGexp}
g_{ij}(r,x)=g_{(0)ij}(x)+r\,g_{(1)ij}(x)+r^2 g_{(2)ij}(x)+\cdots~,
\eea
One substitutes this into Einstein's equations and solves them with the given boundary data. We summarize the main results here (De Haro et al.~(2001: Section 2)), in the absence of matter fields, i.e.~pure gravity:
\begin{itemize}
\item The coefficients in the above expansion, apart from  $g_{(0)}(x)$ and $g_{(d)}(x)$ (the coefficients of the terms with powers $r^0$ and $r^d$, respectively, where $d=D-1$), are all determined algebraically from Einstein's equations. They are given by covariant expressions involving $g_{(0)}$ and $g_{(d)}$ and their derivatives.
\item The coefficients $g_{(0)}(x)$ and $g_{(d)}(x)$  are not determined by Einstein's equations (only the trace and divergence of $g_{(d)}$ are determined): they are initial data.
\item We recover pure AdS (Lorentzian or Euclidean) when $g_{(0)ij}(x)$ is chosen to be flat (i.e. a flat Minkowski or Euclidean metric). In that case all higher coefficients in the series \eq{FGexp} vanish and we are left with \eq{metric}.
\item The case including matter fields (scalars, gauge fields, etc.) can be treated similarly; for details, cf.~Section \ref{matter}.
\end{itemize}
Notice that $g_{(0)}$ and $g_{(d)}$ are, {\em a priori}, arbitrary and unrelated. However, for a specific solution the requirement of regularity of the solution in the deep interior can provide a relation between the two, which is in general non-algebraic.\footnote{This holds for Euclidean signature. For the corresponding statement when the signature is Lorentzian, see Skenderis et al.~(2009: Section 3).} For the sake of calculating quantities, however, it is not necessary to be able to state this relationship explicitly: it is sufficient to assume its existence.

Having given a systematic solution of Einstein's equations that takes into account the boundary conditions, we now have to construct the quantities. The basic quantity in a quantum theory of gravity is the path integral evaluated as a function of the boundary conditions. That is, in Euclidean signature, with $G$ as in \eq{FG}:
\bea\label{genfunct}
Z_{\tn{string}}[g_{(0)}]:=\int_{g_{ij}(0,x)\,\equiv\,g_{(0)ij}(x)}{\cal D}G_{\m\n}~\exp\left(-S[G]\right).
\eea
In the absence of matter fields, this is in fact the basic quantity. All other quantities can be obtained by functional differentiation with respect to the (arbitrary) metric, $g_{(0)ij}(x)$.

In the leading semi-classical approximation (i.e. large $N$ and large 't Hooft coupling), the above is approximated by the (on-shell) supergravity action:
\bea\label{class}
Z_{\tn{string}}[g_{(0)}]\simeq e^{-S_{\tn{class}}[g_{(0)}]}~.
\eea

The Hilbert space structure of AdS/CFT is not known beyond various limits and special cases. But if one is willing to enter a non-rigorous discussion, then a good case can be made that: (i) the  two theories can be cast in the language of states, quantities and dynamics; and (ii) when this is done, they are duals in the sense of Section \ref{equivtthies} (especially comment (1)). We make this case in Section 4.2 of De Haro et al.~(2016). Here we just emphasise that the main conceptual point, as regards (i), is that the gravity partition function (\eq{genfunct} for pure gravity, and  \eq{Zf0} below with matter fields) does not---and should not!---give correlation functions of bulk operators. It gives {\em boundary} correlation functions (of canonical momenta). Accordingly, the evidence in favour of (ii) is largely a matter of  a detailed match (in symmetries; and in quantum corrections to the dynamics as given by a functional integral) between two Hilbert spaces, equipped with operators, both associated with the {\em boundary}.

\subsubsection{The boundary theory and the bulk-boundary relation}\label{bulkbdy}  

Let us now motivate how the bulk geometry is related to the CFT quantities so that we can set up the bulk-to-boundary dictionary. The bulk diffeomorphisms that preserve the form of the line element \eq{FG} modify $g_{(0)ij}(x)$ only by a conformal factor. Thus, the relevant bulk diffeomorphisms are those that induce conformal transformations of the metric on the boundary. 
Since the conformal group  is also the symmetry group of the CFT, it is natural to identify the boundary metric, defined up to conformal transformations, with the classical background metric in the CFT, which is also defined only up to such  transformations. Thus, the correspondence relates $g_{(0)ij}(x)$, up to a conformal factor, with the background metric in the CFT, and the latter need not be flat. We will now use this to set up the AdS/CFT correspondence.



The AdS/CFT correspondence, in the formulation proposed by Witten (1998a) (see also Gubser et al.~(1998)), declares equation \eq{genfunct} to be equal to the generating functional $Z_{\tn{CFT}}[g_{(0)}]$ for (dis-)connected correlation functions, whose logarithm gives (minus) the generating functional for connected correlation functions in the CFT:
\bea\label{log}
W_{\tn{CFT}}[g_{(0)}]:=-\log Z_{\tn{CFT}}[g_{(0)}]~.
\eea
$Z_{\tn{CFT}}[g_{(0)}]$ is the partition function of the theory, defined on an arbitrary background metric $g_{(0)}$. Thus, AdS/CFT is the statement:
\bea\label{adscftpuregrav}
Z_{\tn{string}}[g_{(0)}]\equiv Z_{\tn{CFT}}[g_{(0)}]~.
\eea
This identification is natural because on both sides of the correspondence, and to the order to which the approximation is valid, this is the unique scalar quantity satisfying all the symmetries and depending on $g_{(0)}$, and nothing else. It also makes sense because, as discussed in section \ref{bulk}, $g_{(0)}$ is the asymptotic value of the bulk metric \eq{FG} on the bulk side as well as the classical background metric in the CFT, with their symmetry groups identified (as discussed at the end of Section \ref{adscftintroduced}).

In the semi-classical limit, the left-hand side of \eq{adscftpuregrav} is approximated by the semi-classical action \eq{class}. So using \eq{log} we have the semi-classical correspondence:
\bea\label{sclass}
S_{\tn{class}}[g_{(0)}]\simeq W_{\tn{CFT}}[g_{(0)}]~.
\eea

In practice one wants to calculate, not only  the generating functional for connected correlation functions \eq{sclass}, but also the correlation functions themselves. As in any quantum field theory,  the first functional derivative gives the expectation value of the stress-energy tensor, $\bra T_{ij}(x)\ket_{\tn{CFT}}$, in the CFT with metric $g_{(0)}(x)$:
\bea\label{Tij}
\bra T_{ij}(x)\ket_{\tn{CFT}}={2\over\sqrt{g_{(0)}}}{\d W_{\tn{CFT}}[g_{(0)}]\over\d g^{ij}_{(0)}}~.
\eea

This is related, through \eq{sclass}, to the variation of the bulk quantum effective action with respect to the asymptotic value of the metric, which is the (properly renormalized) gravitational quasi-local Brown-York stress-energy tensor $\Pi_{ij}$ (Brown et al.~(1993)) defined at the conformal boundary. Roughly speaking, the Brown-York stress-energy tensor describes the flux of gravitational energy and momentum at infinity. Using \eq{FGexp} it can be shown that it is given by the coefficient $g_{(d)}(x)$, up to known local terms (De Haro et al.~(2001: Section 3)):
\bea\label{BY}
\Pi_{ij}(x)={2\over\sqrt{g_{(0)}}}{\d S_{\tn{class}}[g_{(0)}]\over\d g^{ij}_{(0)}(x)}= {d \, \ell^{d-1} \over 16\pi\GN}~g_{(d)ij}(x)+ \mbox{(local terms)}~,
\eea
where the local terms involve powers of the curvature. The semi-classical bulk-boundary correspondence for the one-point function is thus: 
\bea\label{1stress}
\Pi_{ij}(x)\equiv \bra T_{ij}(x)\ket_{\tn{CFT}}~.
\eea
For any given solution of Einstein's equations, both quantities are basically given by $g_{(d)}$, which is easily computed from the near-boundary expansion \eq{FGexp}.

We mentioned, in the second bullet point after \eq{FGexp}, that $g_{(d)}$ is the second coefficient that Einstein's equations leave undetermined. The bulk-to-boundary correspondence thus gives two alternative interpretations for $g_{(d)}$ (Balasubramanian et al.~(1999: Section 1)): (i) as the Brown-York stress-energy tensor associated with the boundary, in the bulk theory as in \eq{BY}; (ii) as the 1-point function of the stress-energy tensor in the CFT, through the correspondence \eq{1stress}.

Correlation functions of the boundary stress-energy tensor $\bra T_{ij}(x_1)\,T_{kl}(x_2)\cdots\ket$ can be obtained by taking further functional derivatives in \eq{BY}.\footnote{For higher correlation functions, higher-curvature corrections may contribute terms of the same order in \eq{genfunct} and thus have to be taken into account in the effective action.} From it, the CFT correlation functions can be obtained using \eq{1stress}.

\subsubsection{Matter fields}\label{matter} 

The situation for matter fields is similar. We first consider a scalar field $\f$. Similarly to the bulk metric (cf.~the discussion of \eq{FGexp} in Section \ref{bulk}), one solves the equation of motion perturbatively in the distance $r$ to the boundary:
\bea\label{expandf}
\f(r,x)=r^{\D_-}\f_{(0)}(x)+r^{\D_-+1}\f_{(1)}(x)+\ldots+r^{\D_+}\f_{(2\D_+-d)}(x)+\ldots~,
\eea
where $\D_\pm={{d\over2}\pm\sqrt{{d^2\over4}+m^2\ell^2}}$ and $m$ is the mass of the field. Plugging this into the equations of motion, one finds that all the coefficients $\f_{(1)}(x)$, $\f_{(2)}(x)$, etc., are determined algebraically in terms of the two coefficients $\f_{(0)}(x)$ and $\f_{(2\D_+-d)}(x)$. But these are themselves not determined by the equations: they correspond to boundary conditions. 

As in the case of pure gravity, the first coefficient in the expansion, $\f_{(0)}$, corresponds to a fixed source that couples to a gauge invariant operator ${\cal O}_{\D_+}(x)$ of dimension $\D_+$. The second coefficient, $\f_{(2\D_+-d)}$, then corresponds to the expectation value of that operator $\bra{\cal O}_{\D_+}(x)\ket$ for which $\f_{(0)}$ is a source in the path integral. This can be calculated from the bulk as follows. Take the analogue of \eq{genfunct} for the scalar field\footnote{Here, we are for simplicity taking the metric to be fixed and suppressing it in the notation; also, the final approximate equation is the analogue of \eq{class}.},
\bea\label{Zf0}
Z_{\tn{string}}[\f_{(0)}]=\int_{\f(0,x)\,=\,\f_{(0)}(x)}{\cal D}\f~\exp\left(-S[\f]\right)\simeq \exp\left(-S_{\tn{class}}[\f_{(0)}]\right).
\eea
The AdS/CFT correspondence now declares (cf.~\eq{adscftpuregrav}) that this is equal to the generating functional in the CFT:
\bea\label{adscft}
Z_{\tn{CFT}}[\f_{(0)}]=\bra\exp\left(-\int\dd^dx~\f_{(0)}(x)\,{\cal O}_{\D_+}(x)\right)\ket=:\exp\left(-W_{\tn{CFT}}[\f_{(0)}]\right)
\eea
The expectation value is then calculated by functional differentiation as usual, and it can be shown (using the bulk calculation: i.e.~\eq{expandf} and \eq{Zf0}) that the result is indeed $\f_{(2\D_+-d)}$:
\bea
\bra{\cal O}_{\D_+}(x)\ket_{\f_{(0)}}=-{\d W[{\f_{(0)}}]\over\d {\f_{(0)}}(x)}=(2\D_+-d)~\f_{(2\D_+-d)}(x)~.
\eea
This formalism can be generalised to other kinds of matter fields than a scalar field $\f$. The coupled gravity-matter system, including the back-reaction, can also be treated in the same way (De Haro et al.~(2001: Section 5)).

\subsection{Background-independence}\label{backindpdce}

In the previous subsection we concentrated on the most general form of the AdS/CFT dictionary rather than concrete examples. With this dictionary in hand, we can now discuss the extent to which we have a good theory of quantum gravity. Are the basic quantities, $Z_{\tn{string}}[g_{(0)}]$ and its derivatives (or, if we include matter fields, $Z_{\tn{string}}[g_{(0)},\f_{(0)}]$), the kinds of quantities we expect from a theory of quantum gravity? In particular, the quantities must be invariant under the symmetries of the theory. For a theory of gravity, this is often taken to lead to the important requirement that the theory be `background-independent'. In this Section we will concentrate on this question.

As pointed out in Belot (2011), background-independence is not a precise notion with a fixed meaning. ``[S]peaking very roughly and intuitively, a theory is background-independent if and only if its most perspicuous formulation is generally covariant'' (Belot (2011)). Intuitively speaking, this notion contains two aspects (see also Giulini (2006)): 1) general covariance; 2) absence of `absolute or unphysical structures', i.e.~structures that are themselves not subject to the equations of motion. These two principles will be explicated in (i)-(ii) below, which is called the {\it minimalist conception} of background-independence. Furthermore, there is an {\it extended conception} of background-independence, which will in addition add (iii) below: roughly, the condition that: 3) the boundary conditions also be background-independent.\footnote{An earlier version of this paper only mentioned the minimalist conception of background-independence. In order to clarify the notion, and also in the light of De Haro (2016b) and Read (2016), we have added an exposition of the extended conception of background-independence, from De Haro (2016). For a careful treatment of diffeomorphism invariance, see De Haro (2016a).}

Before proceeding, let us cash out the difference between the minimalist and the extended conceptions of background-independence, for gauge/gravity dualities, in a simple way as follows. The {\it minimalist conception} is the requirement of the background-independence of the {\it bulk theory}, i.e.~the gravity theory in the $(d+1)$-dimensional spacetime: so it is the sense in which general relativity is itself background-independent. The {\it extended conception}, on the other hand, is the requirement that the {\it duality itself} be background-independent, i.e.~that the boundary conditions, which, as seen in Section 6, are the main ingredients of the `dictionary' between the two dual sides, i.e.~constituting what is physical for both sides of the duality, be free from such background-dependence.

As stressed in De Haro (2016:~Section 2.3), the minimalist and extended conceptions of background-independence have different {\it aims}. The minimalist conception is a {\it minimal consistency requirement} of a theory quantum gravity. Because the standard for the minimalist conception of background-independence is {\it low}, this notion is closely modelled on general relativity's own background-independence. The extended conception, on the other hand, does aim at the construction of new theories of quantum gravity, according to some {\it high} standards. Thus this conception is a {\it heuristic principle} for the construction of new theories. Both conceptions, when used in theory construction, are only one out of several principles which one may wish one's theory to satisfy. Thus a quantum theory of gravity should also satisfy usual standards of quantum theories, such as unitarity. In case of an incompatibility between several principles, one may have to weigh the different principles against each other and judiciously assess which is the more important principle to uphold. Another option is, of course, to drop any theory that does not satisfy all of one's principles at once: but usually there is little motivation for such {\it a priori} approach to theory construction. Our understanding of the minimalist and the extended conceptions is thus that the former is the more essential principle to be preserved, while the latter can easily yield to other principles.

\subsubsection{Minimalist vs.~extended background-independence}

More precisely, we maintain, following De Haro (2016: Section 2.3), that {\it minimalist background-independence} should consist of the following two requirements:\footnote{Belot (2011) makes background-independence dependent upon an {\it interpretation} of a theory. Our notion is closer to that of Pooley (2015), but also differs from his, in that we consider quantum theories.}
\begin{itemize}
\item[(i)] There is a generally covariant formulation of the dynamical laws of the theory that does not refer to any background metric, background or unphysical fields,\footnote{An earlier version of this paper did not contain the addition of `unphysical fields'. We thank James Read for pointing out the necessity of explicitly ruling out unphysical fields, e.g.~for cases of some special theories which philosophers have considered. See the discussion in De Haro (2016b) and Read (2016).} `Dynamical laws' is here understood as in Section \ref{dictionary}, i.e.~in terms of an action and a corresponding path integral measure (alternatively, a set of classical equations of motion with systematic quantum corrections); and `background' refers to fields whose values are not determined by corresponding equations of motion. `Unphysical fields' refers to fields whose degrees of freedom are not considered to be physical, by the theory's own lights.
\item[(ii)] The states and the quantities are invariant (or covariant, where appropriate) under diffeomorphisms, and also do not refer to any background metric.
\end{itemize}
The first condition corresponds to the general conception of background-independence as in the works quoted above. The second condition is novel, since those works focus on the equations of motion without defining the physical `quantities' of interest. The rationale for adding a covariance/invariance condition on the states and the quantities, and not only on the dynamics, is the conception of a theory as a triple of states, quantities, and dynamics, that we will introduce in Section \ref{equivtthies}. Roughly speaking, the idea is that a lack of covariance/invariance, or dependence on a background, of the states and the quantities, is a threat to the background-independence of the entire theory: for it would mean that some of the states or quantities would depend on a particular background, or would not satisfy usual standards of covariance. As we will see, gauge/gravity dualities give insight into what those states and quantities in a theory of quantum gravity may be.

As mentioned above, the {\it extended conception} of background-independence adds an additional requirement:

(iii) Any putative initial or boundary conditions needed to solve the theory must also be obtained dynamically from the theory, i.e.~the dynamics of the theory must be such that no externally imposed initial or boundary conditions are required. That is, the same conditions of background-independence which were imposed in (i)-(ii) on the states, quantities, and dynamics of the theory (for gauge/gravity duality: they were imposed on the theory on the {\it gravity side of} the duality), should also be imposed on any boundary conditions that the theory (its states, quantities, or dynamics) may depend on. Furthermore, the theory must be covariant under {\it all} smooth coordinate transformations, including `large' ones.

Since, in the case of gauge/gravity dualities, the boundary values of the bulk fields contain the dynamical information about the dual theory, the requirement of extended background-independence amounts to the requirement that the {\it duality itself} should be background-independent (in the senses (i)-(ii)). 

\subsubsection{Is gauge/gravity duality background-independent on either of the two conceptions?}

We now proceed to argue that, in fact, the conditions (i) and (ii), for {\it minimal} background-independence, {\em are} satisfied for gauge/gravity dualities.\footnote{The ensuing discussion summarises De Haro (2016, section 2.3).} We will briefly discuss condition (iii), for {\it extended} background-independent, in the last paragraph of this subsection.

The first condition, (i), is automatically satisfied because the semi-classical limit of the bulk theory is general relativity (with a negative cosmological constant, and specific matter fields) with a systematic series of quantum corrections: all of which are generally covariant and contain no background fields. 

As to the second condition, (ii), there are two kinds of possible threats:
\begin{itemize}
\item[(a)]dependence of either the states or the quantities on a choice of background metric;
\item[(b)] failure of covariance of the states and the quantities.
\end{itemize}
As for (a): na\"ively, there seems to be an explicit dependence in both the states and the quantities \eq{genfunct} on the boundary condition for the metric, $g_{(0)ij}(x)$. As we saw, the coefficients of the solution \eq{FGexp}, including $g_{(d)}$, depend on $g_{(0)}$. Dependence on a boundary condition, however, is not a breach of {\it minimalist} background-independence: the laws must be invariant (or covariant, where appropriate)  under the symmetries of a theory, but the boundary conditions need not be: applying a diffeomorphism to the boundary condition gives us, in general, a new solution. So one naturally expects that physical quantities {\it will depend on a choice of $g_{(0)}$}. In fact, {\it any} theory whose laws are given by differential equations will require initial or boundary conditions. 
One could speak here of a {\it spontaneous} breaking of the symmetry, in the sense of dependence of the quantities on a particular choice of solution. It is, however, important to realise that this boundary condition is {\it arbitrary}. The boundary conditions thus determine the parameter space of the theory. Thus, point (a) is not a threat to the minimalist conception of background-independence.
This conclusion is unmodified by the addition of matter fields in \eq{Zf0}, for these are covariant as well.

The second threat, (b), is more subtle. It amounts to the question whether the states and the quantities are properly invariant (covariant) under the diffeomorphism symmetries of the theory. Given that the boundary conditions (in particular: $g_{(0)ij}(x)$) are the only parts of the metric that the states and quantities depend on (as per our answer to (a)), this reduces to the question  whether the states and the quantities are properly invariant (covariant) under conformal transformations at infinity. In other words, the question is whether the conformal symmetry is respected. By construction, this must be so away from the boundary. The path integral \eq{genfunct} is invariant under conformal transformations; but only up to divergences coming from the infinite volume of the spacetime (Henningson et al.~(1998: Section 2)). This divergence needs to be regularized and renormalized. But for even $d$, there is no renormalization scheme that preserves all of the bulk diffeomorphisms, and the anomaly cannot be removed.\footnote{On pain of getting a stress-energy tensor that is not conserved, and so breaking a different part of the symmetry algebra: see Deser et al.~(1993).} For odd $d$, on the other hand, there is no anomaly. The anomaly for even $d$ is precisely matched by the anomaly of the CFT when coupled to a curved background $g_{(0)ij}(x)$. The stress-energy tensor, rather than transforming as a tensor, picks up an anomalous term\footnote{For the general expression, see De Haro et al.~(2001:  Section 4). The two-dimensional transformation law is also in Balasubramanian et al.~(1999: Section 3.1).}. This is explicitly seen by taking the trace of equations \eq{BY}-\eq{1stress}; the trace of the right-hand side of \eq{BY} is known and turns out to compute this conformal anomaly. However, this anomaly  does not lead to any inconsistencies of the theory. Also, since the anomaly only depends on $g_{(0)ij}(x)$, this is not a threat to minimalist background-independence; for, as we saw in the case of (a), it is only covariance with respect to the bulk metric---and not the boundary metric---that is needed. 

In conclusion: although for even $d$ the observables depend on the choice of conformal class and so are not diffeomorphism-invariant because of the anomaly; this anomaly has to do with the transformation properties of the metric at infinity, which is fixed (by boundary conditions) and not dynamical. But, on the minimalist conception, there is no reason of principle why the observables should be invariant under these transformations at the quantum level, since the boundary value of the metric is not being integrated over in the path integral. Thus
the theory is also background-dependent in sense (ii).

An interesting further lesson from (b) is that the claim (dating from Kretschmann (1917) and often seen in the literature about background-independence) that covariance is `cheap', i.e.~that it is easily realised in any theory if only the right variables are chosen, is {\it false}. In classical theories with boundaries, as well as in CFT's, there is an anomaly that breaks diffeomorphism invariance and cannot be removed. Thus, we learn from gauge/gravity dualities a general lesson for background-independence: namely, that it is essential to: (i) consider not only equations of motion, but also quantities; (ii) one must specify the relevant class of diffeomorphisms.

Finally, let us discuss whether gauge/gravity duality satisfies condition (iii), which is the additional condition for {\it extended} background-independence. From the above discussion, it is clear that the answer is no, at least for the standard formulations of the duality, because a choice of boundary conditions is required. Though the boundary conditions are arbitrary (and in that sense there is an independence from them), they must be fixed externally (by hand), i.e.~solving the equations requires {\it a choice of} boundary conditions, which is arbitrary and is not made in a dynamical way. Second, in addition to the dependence on the boundary conditions, there is also the {\it lack of covariance} of the quantities, for even values of $d$ (i.e.~for 3-, 5- or 7-dimensional bulk theories), which as discussed above is due to the diffeomorphism anomaly. However, as pointed out in De Haro (2016: Sections 2.3.4 and 3.6) and De Haro (2016b) and Read (2016), there are some cases in which it is possible to effectively integrate the generating functional of the theory, \eq{genfunct} (against appropriate boundary terms) over its set of boundary conditions, so that there is no longer any dependence on them. Through the correspondence \eq{adscftpuregrav}, this then corresponds to coupling the CFT to {\it dynamical gravity}, so that the CFT itself becomes a theory of gravity, with its own conception of background-independence. Thus interpreted, the extended conception of background independence amounts to the background-independence {\it of the duality itself}, as we claimed above. For details, see the one but last paragraph\footnote{The last but one paragraph of Section 2.3.4 of De Haro (2016) wrongly quotes De Haro (2001) for cases of more general boundary conditions. It should read: De Haro (2009)!} of section 2.3.4 in De Haro (2016). As discussed in these references, it is {\it possible} in some cases to couple gauge/gravity dualities to dynamical gravity on the boundary in this way: whether, and it what sense precisely, this should also be seen as {\it necessary}, is still an open problem.

\section{Holography: More General Examples}\label{hologeneral}

As we saw in Section \ref{adscftintroduced}, the AdS/CFT correspondence between $\mathcal{N}=4$ superconformal Yang-Mills and string theory on AdS$_5\times S^5$ is obtained by a careful study of the dynamics of D3-branes in the appropriate limits. More generally: analysing brane dynamics can provide very strong evidence for particular instances of the AdS/CFT correspondence, where the field theory and dual supergravity theory can both be exactly identified.
 For example, another well-studied duality is that arising from a configuration of D1 and D5-branes, where the D1-branes are coincident with (one direction of) the D5-branes. The corresponding CFT lives on this (1+1)-dimensional D1-D5 intersection and is called the D1-D5 CFT; the relevant supergravity solution is AdS$_3\times S^3\times X^4$, where $X^4$ is some compact four-dimensional manifold that the D5-branes are wrapped around (see Maldacena (1998)). 
 We will return to the D1-D5 system again in Section \ref{BHhology}, when we discuss the three-charge D1-D5-P black hole and its dual CFT description.

One can investigate the AdS/CFT correspondence further than these brane-motivated instances. For many practical purposes, it is not really necessary to know the entire content of the dual field theory, but only specific sectors and properties of it. We take this up in Section \ref{bottomup} and Section \ref{hydro}.

\subsection{`Bottom-up' and `top-down' approaches to condensed matter and QCD}\label{bottomup}
One area where the AdS/CFT correspondence is illuminating, even though we do not fully know the dual field theory, is in studies of strongly coupled CFTs that are motivated by condensed matter systems or quantum chromodynamics (QCD).

Certain systems in condensed matter are suspected to be strongly coupled in nature, making their study very difficult using conventional field theory approaches. AdS/CFT provides a way of studying such strongly coupled phenomena. For example, certain high-temperature superconductors are expected to be described by a strongly coupled field theory. Typically, it is not really possible to construct an exact brane-motivated duality for the field theory of a given condensed matter system. However, one can still study the qualitative behaviour of such strongly-coupled field theories, e.g.~superconductivity, by using the AdS/CFT correspondence. One introduces into the bulk theory only the ingredients which are necessary to obtain superconductivity in the field theory: for supergravity, this is simply a gauge field and a charged scalar field. One can remain agnostic about the remaining details and field content of the field theory (and of the supergravity theory) since these will typically not affect the qualitative behaviour.\footnote{Holographic superconductivity was first introduced by Hartnoll, Herzog, Horowitz (2008).}


In such a holographic model of superconductivity, one can distinguish two phases depending on the temperature. For high temperatures, the dominant phase will be a (hairless) charged\footnote{The black hole (hairy or not) must be charged in order to give the dual field theory a finite charge density.} black hole. The charged scalar field is simply zero everywhere in this supergravity solution---this is dual to the field theory's normal (non-superconducting) phase. Below a critical temperature, the supergravity solution that will dominate is that of a charged, ``hairy'' black hole. The hair of the black hole refers to the fact that the charged scalar field now has a non-trivial profile. The non-zero value of the scalar field at infinity is dual to a vacuum expectation value for a scalar field in the dual field theory---the charged scalar condensate that is the order parameter for the superconducting phase. Further properties of these holographic superconducting phases can also be calculated holographically; and they match what one would expect in a superconductor. However, in these holographic superconductors, there do not seem to be any readily identifiable quasiparticles present that are responsible for the conducting properties; this is unlike the standard BCS low-coupling description of superconductivity where these quasiparticles are the Cooper pairs of electrons. Holographic superconductivity may thus provide an interesting viewpoint on the behaviour of recent experimental high-temperature superconductors, which indeed do not seem to be well-captured by the standard BCS theory.


The type of model just described is called `bottom-up', since one simply introduces into the bulk theory the ingredients needed to study the relevant phenomenon in the field theory, even though the exact duality is not known (or needed). In contrast, studies of strongly coupled dynamics using brane-motivated dualities, which give the exact field content of the two dual theories, are called `top-down' models.

Another example of a strongly coupled field theory where both top-down and bottom-up models are used to model it is quantum chromodynamics (QCD). QCD is notoriously hard to study in the regime we are typically interested in, i.e. its confining phase where it confines quarks and gluons to nucleons such as protons and neutrons. This is because in these regimes, QCD is strongly coupled---invalidating pertubation theory approaches. Even though the exact holographic dual to QCD is not known, one can use the AdS/CFT correspondence to study certain aspects 
of strong coupling in theories closely related to QCD. Of course, only qualitative results can be obtained, since we cannot study the exact holographic dual of QCD\footnote{There are a number of reasons why we are not able to study the exact holographic dual of QCD. First of all, QCD is not conformal---although holography has been extended to include some non-conformal field theories as well. An example of one of the more serious problems is that, for QCD, $N=3$, while we saw that quantum corrections are only suppressed in holography if $N\gg 1$.}; but these already provide valuable insights into otherwise calculationally intractable phenomena. For example, confinement and the transition between confinement and de-confinement in strongly coupled gauged theories can be studied with holography.\footnote{Holographic confinement was first noticed by Witten (1998b).} Holographically, this phase transition is the transition between a black hole geometry (de-confining phase) and the so-called \emph{AdS-soliton} (confining phase); one can prove that the AdS-soliton indeed describes a confining phase by calculating the behaviour of Wilson loops in this background.\footnote{A Wilson loop in field theory is the (path-ordered) exponential of the integral of the gauge field around a loop in spacetime. For a field theory to be confining, one expects a Wilson loop to scale with (the exponential of) the area that the loop encloses. A Wilson loop is holographically dual to a string worldsheet area in the bulk that ends on the Wilson loop in the boundary.}


One can try to study other properties of QCD by constructing various models in holography. The top-down model introduced by Sakai and Sugimoto (2005) is a holographic theory involving a strongly-coupled gauge theory with a large number of colors (gluons), with flavors (quarks) introduced. This theory exhibits confinement; also for example, the spectrum of meson masses has been calculated in this theory and been compared to QCD. Another top-down holographic model is the D3-D7 model, first introduced by Karch and Katz (2002), and later expanded upon in a number of ways (see e.g.~Bigazzi, Cotrone, Mas, Mayerson, Tarrio (2012)). This model only has a deconfining phase and is used to study aspects of the quark-gluon plasma. This (deconfined) quark-gluon plasma is exactly the state of QCD matter that is formed at the  Relativistic Heavy Ion Collider (RHIC) and LHC experiments with heavy-ion collisions. It is very difficult to study the quark-gluon plasma with traditional QCD methods as the theory is strongly coupled (even though it is deconfined); holography has provided an opportunity to study aspects of such phases of matter---even though we are not studying the holographic dual to QCD itself. We will return to the quark-gluon plasma and its viscosity as calculated holographically at the end of Section \ref{hydro}.

\subsection{Hydrodynamics and holography}\label{hydro}

In the same spirit as Section \ref{bottomup}'s bottom-up approaches to qualitative properties of strongly-coupled field theories, it is also possible to use the ideas of holography to study the hydrodynamic approximation to a field theory. We begin by recalling the hydrodynamic description of a classical
fluid.

Hydrodynamics is an approximation to the dynamics of a fluid, where one averages over the many microscopic degrees of freedom to obtain a macroscopic description that depends on only a few macroscopic quantities such as the local (timelike) four-velocity $u^{\mu}$, density $\rho$, and pressure $P$. This approximation will only be valid if the fluid is approximately homogeneous, i.e. these macroscopic quantities vary slowly over the fluid.\footnote{To be a bit more precise, the length scale $l_{\tn{hydro}}$ at which we are using hydrodynamics should be much
larger than the mean free path $l_{\tn{mfp}}$ of the fluid particles: so $l_{\tn{hydro}}\gg l_{\tn{mfp}}$.} The hydrodynamic approximation can be improved by including terms with more and more derivatives of $u^{\mu}$; in other words, hydrodynamics is a perturbative expansion in the number of derivatives of $u^{\mu}$. At each level in derivatives, all the possible terms that one can construct with $n$ derivatives introduce new constants. For example, at first order in derivatives, we have for the energy-momentum tensor:
\be 
T^{\mu\nu} = \left[\rho\, u^{\mu} u^{\nu} + P\, (u^{\mu}u^{\nu} + g^{\mu\nu})\right] +\left[ -2\eta\, \sigma^{\mu\nu} - \zeta\, \theta (u^{\mu}u^{\nu}+g^{\mu\nu})\right]    ,
\ee
where $\sigma^{\mu\nu}$ and $\theta$ are particular terms\footnote{For completeness: $\sigma^{\mu\nu} = \nabla^{(\mu}u^{\nu)} + u^{(\mu}a^{\nu)} - \frac13\, \theta(u^{\mu}u^{\nu}+g^{\mu\nu})$,  $\theta = \nabla_{\mu} u^{\mu}$, $ a^{\mu} = u^{\nu}\nabla_{\nu} u^{\mu}.$} containing one derivative of $u^{\mu}$. Thus at zeroth order in derivatives, there are no unspecified coefficients\footnote{However, one does need to supplement the equations of motion with a equation of state that determines $P(\rho)$.} beyond the local functions $u^{\mu}, P, \rho$. But at first order, we see that two new coefficients need to be specified: the \emph{shear viscosity} $\eta$ and the \emph{bulk viscosity} $\zeta$. At second order, we would see an additional fifteen independent coefficients. 

All of these coefficients in the derivative expansion are in principle fixed by the microscopic details of the theory: they are an \emph{input} in the hydrodynamical approximation. But if the underlying microscopical field theory is strongly coupled, it can be practically impossible to calculate, from within the theory, the hydrodynamical coefficients such as the two viscosities. 

However, once again, holography can help in this regard. Using the AdS/CFT correspondence, the hydrodynamical expansion of the energy-momentum tensor $T^{\mu\nu}$ corresponds to an (also derivative) expansion of the (asymptotically) AdS metric $G_{\mu\nu}$ (as in equation (\ref{FG})) on the gravity side. The coefficients of the derivative expansion of $G_{\mu\nu}$ on the gravity are completely fixed by the (classical) equations of motion for the metric; using the correspondence then gives the hydrodynamical coefficients of $T^{\mu\nu}$.

In this way, one can easily prove\footnote{This was originally calculated for $\mathcal{N}=4$ SYM in Policastro, Son, Starinets (2001) and generalized to any CFT with an Einstein gravity dual in Kovtun, Son, Starinets (2005) and Buchel, Liu (2004); reviews of the viscosity/entropy ratio in holography are e.g.~Son, Starinets (2007) and Cremonini (2011).}
that for {\em any} 
conformal field theory dual
to Einstein gravity,
 we have:
\be \label{RHIC} \frac{\eta}{s} = \frac{1}{4\pi},\ee
where $s$ is the entropy density of the fluid. 

This result attracted attention\footnote{See e.g.~Shuryak (2004) for a review from an experimental point of view.} in the experimental world of heavy ion collisions. In these collisions, the quarks and gluons of the ions briefly coalesce in a quark-gluon plasma, which is thought to be a strongly coupled fluid. The experimentally measured ratio of $\eta/s$ of the quark-gluon plasma is extremely low---comparable to $1/(4\pi)$---this cannot be explained using any conventional (weak coupling) description or intuition. Thus the AdS/CFT correspondence provided a tantalizing model in which the shear viscosity of a strongly coupled fluid could be calculated from first principles; moreover, the calculation seems to agree qualitatively with the experimental results about the quark-gluon plasma.

\section{Holographic Theories of de Sitter Space?}\label{dS?}

Current cosmological observations suggest that we live in a universe with an accelerating expansion (see e.g.~Perlmutter (2011)). This means that, to a good approximation, our universe can be described by a spacetime with a positive cosmological constant, an approximation that becomes better and better towards the future, as the universe keeps expanding and the matter content becomes relatively less important. So, to the extent that we are interested in global questions, independent of the detailed matter content of the universe, a spacetime with a positive cosmological constant provides a good model for the future of our universe. 

Just as pure anti-de Sitter space is the maximally symmetric solution of Einstein's equations with a negative cosmological constant (see Section \ref{genpro}, equation \eq{Rmn}), de Sitter space (henceforth: dS) is the maximally symmetric solution of Einstein's equations with a {\it positive} cosmological constant. If a holographic description could carry over to such spacetimes, this would open up the prospect of doing holography for our actual universe! 

One such proposal is the so-called dS/CFT correspondence, where it is assumed that a CFT description of de Sitter spacetime exists. Clearly, dS/CFT would be a much more realistic model of the universe than AdS/CFT; because of its possible cosmological importance, the dS/CFT proposal deserves to be given serious thought.

In Section \ref{difficult}, we will review some of the difficulties that face any theory of gravity in de Sitter space. None of these are specific to dS/CFT; they are  features of the de Sitter space itself, which constitues a time-dependent background. Then in Section \ref{possible}, we will review the progress that has so far been made in solving  these problems, and in developing a holographic description of de Sitter space. 

\subsection{de Sitter spacetime and its difficulties}\label{difficult}

Let us briefly describe the main features of de Sitter spacetime: features by which it resembles its cousin AdS, but with positive cosmological constant. Similarly to AdS in $D$ dimensions (see equation \eq{embed}), dS$_D$ is best introduced via a quadratic equation in a $(D+1)$-dimensional embedding spacetime with one timelike direction:
\bea
-Y_0^2+\sum_{i=1}^DY_i^2=\ell^2~.
\eea
This now resembles the equation for a sphere of radius $\ell$ (positive curvature), rather than a hyperboloid (negative curvature), save for the presence of one timelike direction. 
As in the AdS case (see \eq{sol}), the above embedding equation can be solved by a judicious choice of coordinates, to find the following global coordinates:
\bea
\dd s^2=-\dd t^2+\ell^2\cosh^2(t/\ell)\,\dd\O_{D-1}^2~,
\eea
with $\O_{D-1}$ representing a $(D-1)$-sphere. The topology of this solution is that of a cylinder, with $t$ parametrising the symmetry axis. The spatial volume of the universe is set by the factor $\left(\ell\cosh(t/\ell)\right)^{D-1}$, and there is both a contracting phase and an expanding phase. As time passes from $t\rightarrow-\infty$ to $t=0$, the size of the universe shrinks to a sphere of radius $\ell$; after this phase, the universe expands, its volume growing without bound as $t\rightarrow\infty$. Notice that, unlike AdS space, the spatial slices of dS space are compact for any finite value of $t$: at any given $t$, the spatial geometry is that of a sphere. 

An alternative set of coordinates that covers a local patch between $t=0$ and $t\rightarrow\infty$ is the following:
\bea\label{eta}
\dd s^2={\ell^2\over\eta^2}\left(-\dd\eta^2+\dd{\vec x}^2\right),
\eea
which is the analog of \eq{metric} for AdS. In these coordinates, $\eta=0$ parametrises future timelike infinity (where the volume of the space is infinite), which is now a {\it spacelike} boundary of the manifold. 

We now describe some of the challenges presented by de Sitter space. Again, none of these are specific to a particular holographic proposal: rather, they are intrinsic difficulties about this time-dependent background that {\it any} theory of quantum gravity will have to resolve:
\begin{itemize}
\item The geometry is time-dependent and there is no positive conserved energy (there are no timelike Killing vectors). 

\item The space is expanding, hence each observer has a horizon surrounding them. In other words, because of the rapid expansion, each observer can only see part of the space. This horizon leads to an Unruh effect of particle creation in any state of a quantum field, with a temperature $T={\hbar c\over2\pi k_{\tn{B}}\ell}$.
 Similarly to the case of black holes (see Section \ref{BHhology}), there is an entropy associated with this temperature, and it is equal to one quarter of the horizon area, in Planck units. Unlike black hole entropy, however, this entropy is associated with an observer's {\it cosmological} event horizon surrounding them. The physical interpretation of the microstates associated with this horizon is therefore rather unclear. See e.g.~Spradlin et al.~(2001).

 \item There is no standard theory with unbroken supersymmetry on dS as a background. However, de Sitter space has been found in some supergravity or M-theory constructions. Usually these involve a warped product of de Sitter space with a non-compact `internal' space, or unconventional supergravity theories, which have some kinetic terms with the `wrong' sign or have other unconventional features; see e.g.~Hull (2001), Kallosh (2001), Chamblin et al.~(2001). Also, de Sitter space is not known to be a stable solution of string theory, though it can be realized in string theory as a meta-stable solution: see Kachru et al.~(2003).
 \end{itemize}

\subsection{The possible dS/CFT correspondence}\label{possible}

It has been conjectured that there is a CFT describing quantum gravity in four-dimensional de Sitter space. Because of the causal structure, this CFT should live at timelike future infinity ($\eta=0$ in the coordinates \eq{eta}) rather than at spatial infinity. Clearly, if such a CFT exists, its properties would be quite different from those of the dual in AdS/CFT. We will discuss two related proposals in Sections \ref{analytic} and \ref{higherspin}. 

\subsubsection{Analytic continuation}\label{analytic}

As it turns out, interesting information can be extracted by analytic continuation from Euclidean AdS$_4$. By `analytic continuation', we mean $\ell_{\tn{AdS}}\rightarrow -i\ell_{\tn{dS}}$, which corresponds to $\L\rightarrow-\L$.\footnote{Depending on the coordinates chosen, one also needs to analytically continue one of the coordinates.} Under such analytic continuation, the timelike boundary $r=\e$ of AdS (in coordinates \eq{metric}) gets mapped to the spacelike de Sitter boundary $\eta=\e$ (in coordinates \eq{eta}), where $\e\rightarrow0$ is a cutoff introduced to regulate the large-volume divergences: see the discussion of (ii) (b)  in Section \ref{backindpdce}. Under this proposal, the analytically continued AdS$_4$/CFT$_3$ partition function, $Z_{\tn{CFT}}[\f_{(0)},\e]$ of \eq{adscft}, can be interpreted as the late-time wave-function in de Sitter space:
\bea\label{HH}
\Psi[\vf_{(0)};\e]=\bra\vf_{(0)};\e|0\ket:={\cal A}\left(Z_{\tn{CFT}}[\f_{(0)};\e]\right),
\eea
where ${\cal A}$ is the operation of taking the analytic continuation. The de Sitter time $\eta=\epsilon\rightarrow0$ is close to timelike future infinity, so the result of this is the overlap between: on the one hand, the Bunch-Davies (Hartle-Hawking)\footnote{This is the unique wave-function obtained by analytic continuation from the 4-sphere.} vacuum for de Sitter space $|0\ket$, which acts as a boundary condition in the past; and the final state $\bra\vf_{(0)};\e|$ in the future, where the fields are held fixed. In a similar way to how fields are held fixed at the spatial boundary of AdS, fields are now to be fixed at future timelike infinity.\footnote{\label{f30}There is evidence that one may also need to integrate over these final configurations. If so, the boundary theory will be a theory of gravity, rather than an ordinary CFT.}

The specific connection to Euclidean AdS given here is due to Maldacena (2001, 2011). Thus the current proposal has been called the Hartle-Hawking-Maldacena proposal; though the general dS/CFT proposal was anticipated by Strominger (2001), Witten (2001), and others. 

The dS/CFT proposal is particularly attractive because it means that many of the ideas and techniques available in AdS/CFT can be carried over, with appropriate modifications, to de Sitter space. Furthermore, the prospect that our universe is (approximately) described by a three-dimensional CFT is in itself an interesting one; for, at least in certain regimes, the CFT is expected to be simpler than a theory of quantum gravity. If, as suggested in footnote \ref{f30}, the CFT contains dynamical gravity, then this will be a simpler theory at any rate (conformal, and three-dimensional). More specifically: in the holographic proposal, the absence of Killing vectors and the time dependence (i.e. first bullet point in Section \ref{difficult}) are not a problem, since one is interested in the late-time quantities, including the quasilocal energy as defined there. Also, as regards Section \ref{difficult}'s second bullet point: the CFT may suggest natural interpretations for the microstates associated to the entropy. Thirdly, as to the third bullet point: it may be simpler to try to define the ultraviolet behaviour of a CFT (or, indeed, to directly embed it in string theory or M-theory) than that of quantum gravity in de Sitter space. Needless to say, the proposal is still a conjecture and it is not known whether such a CFT exists. 

\subsubsection{Higher spin theories}\label{higherspin}

There is a related, but in principle very different, proposal for dS$_4$/CFT$_3$ (see e.g.~Anninos et al.~(2011)). The idea is to consider not  just any theory of quantum gravity in de Sitter space, but a specific one: the theory of higher spins developed by Vasiliev and others (Vasiliev (1990, 1999)). This is a theory of {\it massless} higher spins (of any helicity, including those higher than 2) with unbroken symmetry. There are no-go theorems against the existence of such interacting theories in flat space, but it turns out that the no-go theorems do not hold in spaces with positive or negative cosmological constant, where the theories do exist and contain an infinite number of such fields with all possible spins.\footnote{String theories contain a tower of {\it massive} string modes with spins higher than 2. It has thus been conjectured that the higher spin theories correspond to the high-energy limit of string theories (Gross (1988), Vasiliev (2003)), where the string tension and hence the string mass goes to zero.\label{highe}} 

In AdS$_4$/CFT$_3$, there is good evidence that the CFT$_3$ dual of the Vasiliev higher-spin theory is a known field theory, namely the $\mbox{O($N$)}$ vector model. The analytic continuation $\L\rightarrow-\L$ mentioned in Section \ref{analytic} yields $N\rightarrow-N$ on the CFT side (since $N={1\over G_{\tn{N}}\hbar\,\L}$). This latter `analytic continuation' can effectively be realized by changing the group $\mbox{O($N$)}$ to $\mbox{Sp($N$)}$. The proposal is then that the Vasiliev higher-spin theory in the bulk of dS$_4$ (with certain boundary conditions) is dual to the $\mbox{Sp($N$)}$ vector model (or a model with another gauge group, depending on the boundary conditions)  where the fields are anti-commuting. For the `non-minimal' Vasiliev higher spin theory, one encounters (Anninos et al.~(2011, 2013, 2015)) three-dimensional Chern-Simons theory with gauge group $\mbox{U($N$)}$, rather than the vector model. 

The higher-spin proposal seems appealing because it advances concrete CFT's---namely the Sp($N$) and U($N$) models---to correspond to specific choices of theories and boundary conditions in de Sitter space. In these theories, concrete calculations---classical and quantum---can be carried out, and have been carried out, though comparisons with the bulk remain qualitative (see e.g.~Ng et al.~(2013)). The proposal has most of the virtues of the analytic continuation proposal in Section \ref{analytic}, since it is itself the analytic continuation of the higher-spin theories in the context of AdS/CFT. One drawback is that an action principle for the Vasiliev theory is not known; only its equations of motion are known---so the quantities \eq{HH} cannot be calculated independently on the bulk side. The main theoretical difference from the proposal in Section \ref{analytic} is that there is no embedding in string theory or M theory,\footnote{Unless higher-spin theories are indeed related to the high-energy limit of string theory, as in the suggestion quoted in footnote \ref{highe}.} and the bulk theories involved (theories with an infinite number of higher-spin fields) are significantly more involved than classical general relativity. 

\section{Black Holes and Holography}\label{BHhology}

In gravitational physics, black holes have long been enigmatic objects. Almost a century after Schwarzschild's discovery of the first black hole solution in general relativity, their properties are still not completely understood. Holography offers us an interesting new avenue to try to obtain new insights into the true  nature of black holes in string theory. In Section \ref{BHmicro}, we recall the Bekenstein-Hawking formula for black hole entropy, and then review how Strominger and Vafa's counting of a particular black hole's microstates anticipated the AdS/CFT correspondence. Then in Section \ref{kerr}, we sketch some proposals for extending gauge/gravity ideas to other black holes.\footnote{There are many reviews and lecture notes on different aspects of black holes. Black holes and their properties in general relativity are discussed in the excellent lecture notes of Townsend (1997). Some reviews for the hunt for microstate geometries, and the fuzzball proposal (Section \ref{BHmicro}) are Mathur (2008); Bena, Warner (2008); Bena, El-Showk, Vercnocke (2013). Kerr/CFT and other similar approaches (Section \ref{kerr}) are reviewed in Compere (2012).}

\subsection{Black hole entropy and microstates}\label{BHmicro}
One of a black hole's most confusing properties is its entropy. Using a thought experiment that considers the total entropy of a system before and after matter travels into a black hole, Bekenstein (1973) and Hawking (1974, 1975) argued that: (i) a black hole must be assigned an entropy; (ii) this black hole entropy should be the maximum entropy possible for a system; (iii) the amount of entropy that the black hole carries must be proportional to the area $A_{\tn{BH}}$ of its event horizon:
\be \label{BHentropy} S_{\tn{BH}} = \frac{k_B c^3}{\hbar G} \frac{A_{\tn{BH}}}{4},\ee
where we have made the factors $\hbar, G, c, k_B$  explicit to show how all these fundamental constants are combined in this simple, but beautiful and important, formula.

The formula (\ref{BHentropy}) for the black hole entropy is confusing for a number of reasons. First of all, one expects entropy to be an extensive quantity that scales with the volume of a system instead of with its area, since the degrees of freedom of a system are expected to scale with the volume of the system. Thus it seems that (\ref{BHentropy}) already gives us a hint that, fundamentally, the degrees of freedom for gravity live on an area (instead of in a volume)---as would be expected from holographic ideas.
                                                                                                                               
A thermodynamic entropy is traditionally explained by a statistical mechanical counting of possible microstates for the given macrostate. The entropy $S$ of a given system with fixed macroscopic charges
is then  given by counting the number (or phase space volume) $\Omega$ of microstates that this system can be in:                                                                                                                               
\be \label{entropy2} S = k_B \log \Omega.\ee                                                                                                                    
Assigning an entropy to a black hole thus implies that we should consider our description of the black hole as a thermodynamic coarse-graining of some sort of statistical mechanical description. Applying this conclusion to string theory: this implies we should be able to find microstates of a black hole in string theory: so it is natural to wonder what these microstates could possibly look like, and what their properties are.

In this quest, a first important breakthrough was made by Strominger and Vafa (1996). They managed to count the microstates of a  three-charge D1-D5-P (P being a momentum charge) supersymmetric black hole in string theory.
This is not just a particular black hole; it is also special in being {\em extremal}: we will turn to non-extremal black holes in Section \ref{kerr}.\footnote{An extremal black hole is a black hole that, loosely speaking, has the maximal amount of charge possible for a given mass. What this charge exactly is depends on the black hole. In perhaps the simplest example of an electrically charged, spherically symmetric black hole (the Reissner-Nordstrom black hole), this means that in geometric units, the mass is equal to the charge, $M=Q$. Of course, it is not a coincidence that this ``maximal charge'' condition is often precisely the BPS bound defining supersymmetric states.}
In holographic terminology, Strominger and Vafa were able to perform a combinatoric counting of the relevant CFT microstates that should correspond to the black hole in the dual gravity theory. In other words, they could identify the microstates of the black hole exactly---at least in the CFT description dual to the black hole. Then, their combinatoric CFT computation gave them the microscopic entropy (\ref{entropy2}), which matched beautifully with the known result for the entropy (\ref{BHentropy}) of the three-charge black hole in the dual supergravity theory.

Interestingly, Strominger and Vafa (1996) performed their holographic microstate counting before the advent of Maldacena (1998) and his introduction of AdS/CFT holography. They were influenced by earlier ideas by e.g.~'t Hooft (1990) about interpreting black holes in string theory. Thus various works, including 't Hooft (1990) and Strominger and Vafa (1996), set the stage for the Maldacena's AdS/CFT proposal. In hindsight, the Strominger-Vafa calculation can be seen as a prime example of the AdS/CFT correspondence, where the CFT living on the D1-D5 branes is identified with the asymptotically AdS$_3$ black hole solution of the D1-D5 branes.

While it was possible in this and other specific examples to identify (or at least count) microstates explicitly in the dual CFT, this does not mean that these microstates are readily identifiable in the dual gravity description. In principle, supergravity black hole microstates should be horizonless (so that they each carry zero entropy) and smooth (since they should be well-behaved solutions). However, it is  not at all clear that generic microstates of a black hole will be smooth or otherwise well-behaved in the supergravity approximation. It could be that the microstate geometries are inherently stringy in nature, so that a good description requires the intricate machinery of the full string theory. The search for (well-behaved) supergravity microstates and for ways of counting them is an active area of research and often goes under the name of the `fuzzball proposal'.

\subsection{Kerr/CFT and non-extremal black holes}\label{kerr}
The three-charge black hole considered in the Strominger-Vafa calculation (described in Section \ref{BHmicro}) is asymptotically flat. However, holography is possible because we can zoom in to the near-horizon region of the black hole and throw away the asymptotically flat part of the solution, leaving us with an asymptotically AdS$_3\times S^3\times X^4$ spacetime (where $X^4$ is a compact four-dimensional manifold). This `zooming-in' procedure is completely analogous to the decoupling limit described in Section \ref{adscftintroduced} for D3-branes and AdS$_5\times S^5$: it is essentially a low-energy limit, where the excitations far away from the branes decouple from the excitations near the branes.

The fact that such a decoupling limit is possible is a feature that is shared by many other extremal black holes where we have a clear picture of how the black hole arises from (various species of) branes. In most such cases, the field theory on the branes in question is then dual to the AdS space that arises in the near-horizon limit of the black hole geometry. This readily provides an interpretation and description in the dual CFT of the extremal black hole considered. However, we emphasise that there is no universally accepted way to provide a CFT dual to a general extremal black hole, especially when no brane picture is readily available. The Kerr solution (or Kerr-Newman, when charges are turned on as well) is a solution describing a rotating black hole. The angular momentum of this black hole can be taken to be maximal to get an \emph{extremal} solution. However, even for this extremal Kerr(-Newman) solution, there is no immediate brane description available.
 Accordingly, `the Kerr/CFT correspondence' is a generic name for attempts to provide CFT dualities for larger, general classes of black holes.

Furthermore, extremal black holes are not very realistic. Generic, non-extremal black holes are expected to radiate thermally in the well-known phenomenon of Hawking radiation, by which we can associate a temperature to the black hole. However, extremal black holes have zero Hawking temperature and are thus not expected to radiate. If a black hole is very close to being extremal, the non-extremality can be viewed as a small perturbation; this translates into a slight deformation of the decoupling-limit AdS space. However, for a fully non-extremal black hole, it becomes impossible to perform any meaningful decoupling limit; the excitations near the black hole become inexorably entangled with the excitations far away.

Nevertheless, a few meaningful hints pointing towards holographic descriptions for these black holes also exist. In the program of \emph{hidden conformal symmetry} (introduced in Castro et al.~(2010)), it was realized that general, non-extremal rotating black holes exhibit $\mbox{sl}(2)\times \mbox{sl}(2)$ symmetry in a certain low-energy limit of the equation of motion for a probe scalar field moving on the black hole solution as background. Such a symmetry is reminiscent of CFTs, but is globally broken to $\mbox{U}(1)\times \mbox{U}(1)$, making the symmetry `hidden'.

A separate but related approach is that of \emph{subtracted geometries}. Cvetic and Larsen (2011) argued that the thermodynamics of a general, non-extremal black hole in 4D or 5D is independent of a particular warp factor in the metric that controls its (asymptotically flat) asymptotics. Thus, one can imagine changing this warp factor by hand to have different asymptotics, without actually modifying the black hole itself. Cvetic and Larsen considered a particular choice for this warp factor that allowed an immediate uplift to a higher-dimensional AdS$_3\times S^d$ spacetime (with $d = 2, 3$ depending on the initial black hole), thus immediately suggesting a dual CFT description of the black hole. It is not entirely clear how relevant this CFT description can be for the original, asymptotically flat black hole---see e.g.~Baggio, de Boer, Jottar, Mayerson (2013).

These approaches are certainly very suggestive, but it should be emphasized again that no convincing suggestions exist for how to treat general non-extremal black holes holographically. Suffice it to say that, if holography is indeed valid for all gravitational theories, there should always exist a theory without gravity in which the black hole, its dynamics, and its microstates are completely and equivalently described. However, this dual theory may in general be a very complicated theory that bears almost no resemblance to the  CFT descriptions of extremal black holes that are currently well understood.  

\section{Philosophical Aspects}\label{philaspects}

This Section discusses how gauge-gravity duality bears on two philosophical topics: (i) how physical theories should be individuated, and thus the conditions under which two apparently different theories are really the same (Section \ref{equivtthies}); (ii) the conditions under which one theory, or a feature of the world described by a theory,  can be said to be emergent  (Section \ref{emergence}).\footnote{\label{philrefs}{Our views on these topics are given in more detail in: De Haro (2016), De Haro et al.~(2016) for (i); and De Haro (2016), Dieks et al.~(2015) for (ii).}}

\subsection{The equivalence  of theories}\label{equivtthies}
From Section \ref{intro} onwards, we have described our versions of gauge/gravity duality as involving what we called `equivalent theories'.  And in Section \ref{dictionary}, this was made a bit more precise. Namely, we took a theory to be given by a state-space, equipped with various structures, especially a set of quantities and a dynamics; and we took a duality to be a bijective structure-preserving mapping between theories thus understood. The idea was thus that two theories being equivalent is a matter of their `saying the same thing', as shown by the duality mapping. See, in particular, the last paragraph of Section \ref{bulk}.

We will now briefly develop this construal of duality, with five comments: each leads in to the next. But the first two are general, while the others are specific to gauge-gravity duality (though some remarks clearly apply equally to other dualities in string theory). For all five comments, two points apply: (i) we will assume the duality holds exactly (a substantive assumption, since the dualities we have reviewed are unproven!), so that whether the dual theories are equivalent is indeed an issue; (ii) there are more details in the references cited in footnote \ref{philrefs}. 

(1): {\em Precision}:--- This construal of duality can be made precise, with simple formal definitions of what a theory, and a duality, are. We define a theory as a triple, $\bra{\cal S},{\cal Q},D\ket$,  consisting of: a state-space ${\cal S}$ (classically, some sort of phase space; and in a quantum theory, a Hilbert space); a set of quantities ${\cal Q}$ (classically, some class of real-valued functions on phase space, and in a quantum theory, typically some class of self-adjoint operators); and a dynamics $D$, which we will express in the `Schr\"odinger-picture', i.e. as a time-evolution of states.  States and quantities are assignments of values to each other. So there is a natural pairing, and we write $\langle Q ; s \rangle$ for the value of $Q$ in $s$. In classical physics, we think of this as the system's intrinsic possessed value for $Q$, when in $s$; in quantum physics, we think of it as the (orthodox, Born-rule) expectation value of $Q$, for the system in $s$. Agreed: one could  add other components to this construal of `theory', additional to these three: such as a set of symmetries (of various kinds: dynamical, gauge), and a set of parameters. But we will not need to do so.

We then take a duality to be an {\it isomorphism}, in the obvious sense, between  two theories $\bra{\cal S}_1,{\cal Q}_1,D_1 \ket$ and $\bra{\cal S}_2,{\cal Q}_2,D_2 \ket$. By this, we mean bijections $d_s: {\cal S}_1 \rightarrow {\cal S}_2$ and $d_q: {\cal Q}_1 \rightarrow {\cal Q}_2$, between the theories' sets of states, and sets of quantities, respectively, that ``mesh'' appropriately with: (i) the assignment of values  $\langle Q ; s \rangle$ ; and (ii) the dynamics. Thus for (i), we require, in an obvious notation:
\be\label{obv1}
\langle Q_1; s_1 \rangle_1 = \langle d_q(Q_1) ; d_s(s_1) \rangle_2 \; , \;\; \forall~Q_1 \in {{\cal Q}_1},~s_1 \in {{\cal S}_1}. 
\ee
For  (ii), we require that $d_s$ commutes with (is equivariant for) the two theories' dynamics; i.e. the group actions $D^{\tn{Sch}_i}$ of $\mathR$ on ${\cal S}_i$:
\be\label{obv2}
d_s(s_1(t)) \equiv  d_s(D^{\tn{Sch}_1}(t,s_1)) = D^{\tn{Sch}_2}(t,d_s(s_1)) \; , \;\; \forall~t \in \mathR,~s_1 \in {{\cal S}_1}. 
\ee
Eq. (\ref{obv1}) and (\ref{obv2})  appear to be asymmetric between $T_1$ and $T_2$. But in fact they are not, thanks to the maps $d$ being bijections. 

Thus we have adopted the definition of `duality' that is obvious and simple, given our conception of `theory'. One  could strengthen the definition in various ways: for example, to require that $d_s$ be unitary for quantum theories etc.  And of course, there is a whole tradition of results relating the requirement of matching values \eq{obv1} to such strengthenings: the obvious one in quantum physics being Wigner's theorem that the map's preserving all the transition probabilities implies its being  unitary or anti-unitary. But we do not need to pursue such strengthenings.

(2): {\em Beware: `disjoint but isomorphic'}:--- At this point, a warning is needed. Formal isomorphisms of the kind just discussed cannot {\em guarantee} that two theories are equivalent in the sense we have intended: that they `say the same thing', in the sense of asserting (a) the very same properties and relations about (b) the very same subject-matter.  For it seems that the world could contain two distinct, indeed non-overlapping, subject-matters---or in other words: sectors or realms---each of whose constituent objects have distinctive properties and relations, that are {\em not} instantiated in the other subject-matter---and yet the {\em pattern of instantiation} of the two sets of properties and relations could be formally the same, so that there is a duality isomorphism of the kind defined in (1). That is: although the subject-matters are utterly different,  two theories of the two subject-matters could  be duals---even while each theory is wholly true about its subject-matter. 

To sum up: `saying the same thing' , as in (a) and (b), is a matter of semantics, of what words refer to. And what a word refers to is settled by various (usually complex) conventions rooted in the contingent history of how the word was first used, and how that usage developed---matters which are very unlikely to be captured by formal isomorphisms of the kind defined in (1)! 

Agreed:  we can still claim that theories that are duals in the sense of (1) {\em can} be equivalent: they can `say the same thing'. More positively: by explicitly appealing to the notion of reference, we can characterize {\em when} they will do so. Namely: they do so---they are equivalent---provided the conventional reference in each theory of one of its words $w$ is the same `hunk of reality' as the conventional reference in the other theory of the word $w'$ that corresponds to $w$ in the duality mapping (the `dictionary'). 

So much for generalities. We now discuss how gauge-gravity duality illustrates the themes in (1) and (2): comment (3) discusses the definitions in (1); comments (4) and (5) discuss the warning in (2).

(3): {\em AdS/CFT exemplifies the definition of `duality'}:--- We maintain that AdS/CFT, as summed up in Section \ref{dictionary}'s \eq{genfunct} and \eq{adscftpuregrav} for pure gravity, and in \eq{Zf0} and \eq{adscft} for matter fields, {\em does} exemplify comment (1)'s abstract definition of `duality'. 

To justify this claim would obviously require some work: for Section \ref{dictionary}'s bulk and boundary theories are not given to us with a precisely defined state-space, set of quantities, or dynamics. It would also obviously require some flexibility---some generosity!---about what counts as a justification in the present state of knowledge. For at present, no one knows how to rigorously define these theories: not just the string theories on the bulk side, but also the conformal field theories on the boundary side. For example, to rigorously define the generating functional in a quantum field theory, as in the r.h.s. of \eq{adscft},  requires knowledge of the full non-perturbative structure of the theory, which is at present available only for very few quantum field theories (mainly: topological QFT's, and field theories in two dimensions). 

(4): {\em The consensus despite the warning}:--- The consensus in the string theory community is that, despite our warning in (2), and the disparity of the {\em apparent} subject-matters of the two theories in AdS/CFT (Sections \ref{adscftintroduced} and \ref{dict})---e.g.~their differing about the dimension of spacetime, and whether it is curved---the two theories {\em are} equivalent. A bit more precisely (and---to repeat---assuming that the AdS/CFT duality holds exactly): the consensus  is that the two theories `make exactly the same claims'---not just about observational matters but also about unobservable, i.e.~theoretical, matters. As the physics jargon has it: they `describe the same physics'.\footnote{Besides, this consensus is held not only for AdS/CFT, and the more speculative gauge-gravity dualities of Sections \ref{hologeneral} to \ref{BHhology}: but also for other remarkable dualities in string theory, such as T-duality. As we mentioned in Section \ref{stringrecap}, this is a duality in which the two theories apparently differ about the radius of a compact dimension of spacetime. Roughly speaking: where one theory says the radius is $R$, the other says it is $1/R$: a recent philosophical reference is Huggett (2016, Sections 2.1, 2.2).}

And many---indeed, we think: most---philosophical commentators endorse this consensus, including for our case of gauge/gravity duality; e.g.~Dawid (2007, Section 6.2), De Haro (2016, Section 2.4), Dieks et al.~(2015,  Section 3.3.2), Huggett (2016, Section 2.1, 2.2), Matsubara (2013), Rickles (2011, Section 2.3, 5.3) and Rickles (2015). 

(5):  {\em Theories of the universe, and the internal point of view}:--- There is an obvious explanation for why string theorists disregard the warning in (2), and adopt the consensus reported in (4). For string theory is often taken to be aiming to provide a `theory  of everything', i.e. a theory  of the whole universe. And in that kind of scientific enterprise, the idea of distinct but isomorphic subject-matters tends to fall by the wayside. For such a theory, there will be, {\em ex hypothesi}, only one subject-matter to be described, viz. the universe. Besides, the theory will probably also aim to state enough about the relations between the many sub-systems of the universe, so that no two sub-systems will get exactly the same description; so that even at the level of sub-systems, there will not be distinct but isomorphic subject-matters.\footnote{We should also note another aspect of the consensus reported in (4). Many string theorists hope that pressing the idea that the two dual theories are equivalent, despite their striking differences, will help them to formulate a better theory (the much-sought M theory!) that will underly the present theories in something like the way a gauge-invariant formulation of a theory underlies its various gauge-fixed formulations. For a more detailed comparison of dualities and gauge symmetries, cf.~De Haro et al.~(2016).}
 
Finally, we note that in the context of seeking a theory  of the whole universe, one might go further than setting aside the possibility of distinct but isomorphic subject-matters. One might also hold that the interpretation of our words, i.e. of the symbols in the  theory, must be fixed `from within the theory'. And this last phrase is taken to imply that the interpretation  will be {\em the same} on the two sides of a duality: i.e. the same for a word $w$ in the theory on one side, and for the word $w'$ that corresponds to $w$ according to the `dictionary'. This view is endorsed, under the label `internal point of view', in De Haro (2016, Section 2.4) and Dieks et al.~(2015, Section 3.3.2).
   
\subsection{Spacetime and gravity as emergent?}\label{emergence}

We have seen that gauge/gravity dualities relate theories with completely different properties: in the case of AdS/CFT, a higher-dimensional theory of gravity is related to a lower dimensional quantum field theory. This disparity prompts one to ask two questions:\\
\indent (a) Does the additional spatial dimension in the higher-dimensional theory somehow `emerge' from the low-dimensional theory?\\
\indent (b) Does the force of gravity itself, not present in the low-dimensional theory, `emerge' in the high-dimensional theory? \\
Indeed, claims of `emergent spacetime' and `emergent gravity' abound in the literature about gauge/gravity dualities. What should we make of these claims? 

In this Section, we will: argue that for emergence, the duality concerned can only be {\em approximate} (Section \ref{emapprox}); and then discuss two ways in which, using approximations, spacetime and-or gravity could be emergent (Section \ref{2ways}).

\subsubsection{Emergence vs. duality}\label{emapprox} 

Of course, `emergence' does not have a very precise meaning, even in technical philosophy. But the main idea is that a theory, or a `level of description', is emergent from another if it has features that: (i) are not readily, or perhaps not even in principle, deducible or explainable from the other theory or level; but nevertheless (ii) are somehow grounded, or rooted, in the other theory.  Similarly if we construe emergence in terms of ontology, rather than human descriptions: e.g.~classes of phenomena, or of properties or behaviour, that a system or set of systems exhibits.\footnote{Bedau and Humphreys (2008) is a fine anthology about emergence.} 

In this Section, we will construe emergence along these lines, though a little more precisely, following Butterfield (2011, Section 1.1; 2011a, Sections 1, 3). Namely: an emergent theory describes ``properties or behaviour of a system which are novel and robust relative to some appropriate comparison class''. `Novel' here means ``not deducible from the comparison class'', and ``showing features (maybe striking ones) absent from the comparison class''. And `robust' means that the emergent properties or behaviour are not destroyed if small changes are made in the comparison class. (Here, `smallness' is to be measured in terms of the comparison class: so something like a quantitative measure needs to be provided.) As we will see, dualities provide relevant comparison classes. 

This conception of emergence, like the initial vaguer idea of emergence, is obviously asymmetric: if theory B emerges from theory A, then theory A cannot emerge from theory B. (It is also non-reflexive: A cannot emerge from A, since there can be no novelty and robustness of behaviour as described by a theory, in comparison to itself.) On the other hand, duality is a symmetric relation: if A is dual to B, then B is dual to A.\footnote{This symmetry is a feature of the usual usage in physics, in particular in gauge/gravity duality, as we saw in the previous Sections: as well as, of course, our definition of duality in Section \ref{equivtthies}. And most usage also supports duality being a reflexive and transitive, and so an equivalence, relation: as is implied by our definition.}
  
It thus follows from the mere logic of the terms---not just that the presence of duality is insufficient for emergence---but that on the contrary: duality in fact precludes emergence. Imagine, for instance, that we accept that string theory in AdS$_5\times S^5$ is dual to ${\cal N}=4$ SYM theory on ${\mathbb R}^4$; and we want to claim that the additional radial direction\footnote{Alongside the $S^5$.} within the AdS$_5$ emerges from the SYM theory. But then the duality prevents us from considering gravity and the additional spatial direction as novel properties. For it implies that those properties correspond to other, well-defined, properties on the CFT side, and thus no novel behaviour can arise: there is merely a reformulation of the properties or behaviour. Furthermore, the purportedly novel behaviour could not be robust: for the properties of the new space that appears depend very sensitively  on the dual properties in the CFT (which provides the comparison class). Thus, duality precludes emergence. 

What, then, is one to make of the literature's claims of `emergent spacetime' and `emergent gravity'? This discussion  indicates that we need to somehow `break' the duality in order to be left with an asymmetric relation. This is also suggested by the concept of robustness: we need to introduce some contrast between `macroscopic' and `microscopic' theories; or (better, because scale-independent) between `coarse-grained' and `fine-grained' theories. Then we can hope to capture both the asymmetry and the robustness in terms of the coarse-grained theory being independent of the fine-grained details. 

Such asymmetric relationships {\em do} obtain when we consider particular approximations of the theories that are duals: the approximative relation will introduce a notion of coarse-graining. For instance, in the low-energy limit of the AdS$_5\times S^5$ string theory discussed in Sections \ref{genpro}, \ref{adscftintroduced} and \ref{dict} we get structures, such as the classical AdS$_5$ geometry, that did not figure in the high-energy theory (where there was only an {\it asymptotically} AdS$_5$ space) and can only be defined within a particular approximation (involving large $N$). These properties are thus: (i) novel, i.e.~not defined in the high-energy (fine-grained) string theory; (ii) robust, i.e.~stable against small changes in the high-energy theory. 

\subsubsection{Two ways for spacetime or gravity to be emergent}\label{2ways}

Motivated by this example, we can now distinguish two ways in which spacetime and-or gravity might emerge, in the framework of dualities. For each way, we will first focus on the emergence of theories, and then on the emergence of properties and  behaviour. This distinction is developed in more detail in De Haro (2016, especially Section 3.2; cf.~also Dieks et al.~(2015: Sections 3.3.2, 3.3.3, 4.2 and 4.3)). Once we have the distinction in hand, we will be able to sketch answers to questions  (a) and (b) above.  \\
\\
(i) {\it Emergence across a duality}:--- Here, the idea is that one theory is dual to another, which itself has a coarse-graining scheme, based on some parameter indicating scale (especially length or, inversely, energy). So it is at most for {\em one} level of fine-graining in the second theory, that the duality is exact; and maybe the duality is not exact at any level. In such a scenario, features of the first theory can be novel relative to those of the second: either because the duality is approximate at all levels (indeed, perhaps with very different duality maps); or because although exact at one level, we have reason to consider the second theory  at {\em another} level, where the duality is approximate, so that the features of the first theory show novelty. Besides, these features can be robust in the sense of being independent of various fine-grained details of the second theory. 

Applied to gauge/gravity duality, the idea is thus that a theory G (for `gravity') that is is higher-dimensional and contains gravity, is dual to a theory F (for `fundamental') that is low-dimensional and contains no gravity---but has a coarse-graining scheme. Thanks to the coarse-graining scheme, the duality is exact for at most at one level of fine-graining of F; and so, as explained above, there is scope for G to be emergent from F.  

The `bottom-up' approaches of Section \ref{hologeneral} may exemplify this idea. That is: they may furnish examples of  general relativity (theory G) emerging in this way from condensed matter theory (theory F). There is no exact duality, but the spacetime picture appearing is novel and robust: it is independent of the details of the condensed matter theory. For instance, it was remarked, about \eq{RHIC}, that the viscosity-to-entropy ratio is independent of the CFT under consideration. And it is the approximate duality itself that provides the relevant comparison class required for robustness: a family of CFT's at different values of the couplings.\\
\\
(ii) {\it Emergence on both sides of a duality}:--- Here, the idea is that each side of the duality has a coarse-graining scheme, so that there can be emergence {\em independently} on the two sides. And in principle, levels on one side could be paired by duality with levels on the other: in which case, there would be a sequences of dualities parametrised by the pairs of levels being mapped to each other.

Applied to gauge/gravity duality, this means that theory G emerges from theory G'  via an approximative relation that leads to the appearance of novel and robust spacetime structures; and on the other (boundary) side of the duality, theory F emerges from theory F'. So there are parallel cases of emergence on both sides: F and G emerge out of F' and G', respectively. The emergent behaviour is  independent of the duality (or dualities): there would be emergence even if there were no duality. The duality (or dualities) imply only that we get two emergent behaviours---and perhaps with very different appearances and structures  on the two sides.

For instance, the classical AdS spacetime geometry (in theory G: general relativity) emerges from the situation in which there is no classical spacetime geometry (in the fine-grained string theory: theory G'), but only the fundamental quantum fields living on the string. In Dieks et al.~(2015, Sections 3.3.3, 4.2) the `renormalization group flow' was identified as a coarse-graining mechanism through which a particular energy regime of a quantum field theory is explored. And it was argued that in this regime, emergent behaviour can arise.\footnote{Similar points apply to a conventional quantum field theory (Butterfield (2014: Section 4.3); Bouatta and Butterfield (2015a: Sections 3.1, 4.2)): the dwindling contribution of the non-renormalizable terms, as the renormalization group flows to the infra-red, makes renormalizability emergent.} In the bulk there is a similar process through which the radial direction can be seen to be emergent.  Such limits have also been considered in Teh (2013, Section 3.3, 4).\\
\\
In De Haro~(2016: Section 3.2) it was argued that these two forms of emergence:\\
\indent (A) correspond to two different ways in which dualities can be broken by coarse-graining: and the two ways correspond to different properties of the duality isomorphism; \\
\indent (B) exhaust the possibilities for emergence from dualities via coarse-graining.\\
Furthermore: because emergence is now associated with {\it approximate duality} rather than with specific gauge/gravity dualities, these two forms of emergence may apply to other dualities. 

With the distinction between these two forms of emergence, (i) and (ii), in hand, let us now go back to our original questions (a) and (b).\\

 Regarding (b), the emergence of gravity:--- By construction, gravity emerges in both kinds of emergence. For {\it general relativity} (appropriately coupled to matter) is shown to emerge in both (i) and (ii); hence, with it, the gravitational force. This is straightforward in (i), where the fundamental theory F is clearly not a theory of gravity but, say, a quantum field theory. In (ii), the verdict is more subtle: for G now emerges from the fine-grained theory G' (say, string theory); and the latter might, at the fundamental level either (1) be, or (2) not be, a theory of gravity. And the answer to that question, of course, depends on a fundamental formulation of string theory, which is still lacking. (1) If string theory is at the fundamental level {\em not} a theory of gravity, then one must of course explicate in what sense it is not, i.e.~one needs a first-principles account of what counts as a theory of `gravity'. But that gravity is indeed emergent is then straightforward. (2) On the other hand: if string theory {\it is} a theory of gravity at the fundamental level, then it is only {\it Einsteinian}  gravity that emerges after coarse-graining. Either way, gravity (in the sense of a specific law of gravity) emerges in an uncontroversial sense.

Regarding (a), the emergence of one dimension of space:--- In (i), this is uncontroversial, since theory G has one more dimension than theory F; and G emerges from F {\em a la} (i). In (ii), the verdict is again subtle: for G does not emerge from F but from G'; and the number of dimensions in G and G' is the same. Thus, in (ii) there is no emergence of a dimension of space in this sense; (though there could be emergence of {\it all} space, and time, dimensions: namely if the fundamental theory G' is not a spatio-temporal theory to begin with). Space thus only emerges uncontroversially within the first scheme, (i).

Finally, a remark about the emergence of spacetime and gravity, in the context of cosmology. Recall from the start of Section \ref{possible} that in de Sitter space, the holographic  direction (i.e. the radial direction, with gravity) is timelike rather than spacelike, as is seen e.g.~in \eq{eta}. Thus, if dS/CFT exists and emergence is realised in  sense (i) above: then for dS/CFT, it is {\it time}, rather than space, that is emergent. And in {\it both} (i) and (ii), Einsteinian gravity emerges as time moves forward. Indeed, Strominger (2001: Abstract) has argued that ``the monotonic decrease of the Hubble parameter corresponds to the irreversibility of renormalization group flow.''  Thus, both in (i) and in (ii), time evolution is the coarse-graining process that leads to emergent structures.

\section{Conclusion}\label{concl}

In this paper we have surveyed the `landscape' of gauge/gravity duality: the key idea being that a theory  of gravity in $D$ dimensions is dual to a theory  without gravity in $D - 1$ dimensions. At the centre of the landscape is the AdS/CFT duality, reviewed in Sections \ref{adscftintroduced} and \ref{dict}. The ingredients of this duality---AdS spacetime, conformal field theories, and string theory---were reviewed in Sections \ref{genpro} to \ref{stringrecap}. Then we reviewed various other proposals for gauge/gravity dualities, including proposals for condensed matter theory and other real-world strongly coupled field theories (Section \ref{hologeneral}), de Sitter spacetime (Section \ref{dS?}) and black holes (Section \ref{BHhology}). Finally in Section \ref{philaspects}, we discussed the general ideas of (i) two theories being equivalent to each other, and (ii) spacetime or gravity as described by one theory being emergent from another.

The wide range of this discussion, and the fact that countless research problems are still open, makes it clear that it would be very premature to attempt any systematic conclusions. So we end simply, by emphasising some themes that seem to us central to the conceptual aspects of the subject; as follows.
\begin{itemize}
\item[(1)] Background-independence in AdS/CFT (Section \ref{backindpdce}):--- AdS/CFT is background-independent in a suitably minimalist sense. The minimalist sense is strong enough for AdS/CFT to qualify as a candidate theory of quantum gravity when the cosmological constant is negative. But, against the received view going back to Kretschmann, general covariance is sometimes broken by anomalies even in classical theories; and when discussing background-independence one must be careful to identify the {\it relevant} class of diffeomorphisms.
\item[(2)] The problems and prospects for de Sitter holography (Section \ref{possible}):--- The spacelike nature of the holographic boundary raises interesting philosophical questions about the nature of the physical quantities  (sometimes called `meta-observables') that figure in the dS/CFT proposal, since all of these quantities are not accessible to any single observer. This seems to provide further physical reason to distrust the traditional positivist inference from  unverifiability to meaninglessness. 
\item[(3)] The problems and prospects for black hole holography, especially for non-extremal cases (Section \ref{kerr}): The puzzles involving black holes (concerning their entropy and microstates, and the information paradox) may be clarified or understood better by holographic methods, even though it remains difficult to provide descriptions for non-extremal black holes.
\item[(4)] The relations between emergence and duality (Section \ref{emergence}):--- Despite the fact that the presence of a duality between two theories {\it prima facie} precludes emergence of one theory from the other, duality combined with coarse-graining can make up for emergence. Emergence can occur in two very different  ways, according to two different ways of applying coarse-graining.
\end{itemize}


\section*{Acknowledgements}
\addcontentsline{toc}{section}{Acknowledgements}

JB and SdH thank Nicholas Teh for discussions. SdH also thanks the audiences and organisers of the conferences `Equivalent Theories in Physics and Metaphysics' at Princeton University and `Emergent Time and Emergent Space in Quantum Gravity' at the Max Planck Institute for Gravitational Physics. 
DRM was supported in part by a Cottrell Scholar Award from the Research Corporation for Science Advancement. This work is part of the research programme of the Foundation for Fundamental Research on Matter (FOM), which is part of the Netherlands Organisation for Scientific Research (NWO).

\section*{References}
\addcontentsline{toc}{section}{References}

Aharony, O.,  S.S. Gubser, J.M. Maldacena, H. Ooguri, and Y. Oz. (2000). ``Large \emph{N} field theories, string theory and gravity'',  \emph{Physics Reports}, 323(3-4), 183-386.  [hep-th/9905111].\\
\\
Ammon, M., Erdmenger, J.~(2015). {\it Gauge/Gravity Duality. Foundations and Applications}. Cambridge University Press, Cambridge.\\
\\
Anderson, J.~(1964). ``Relativity principles and the role of coordinates in physics.'' In: H.Y. Chiu, W. Hoffman (eds.) {\it Gravitation and Relativity}, pp. 175-194. W.A. Benjamin, Inc., New York.\\
\\
Anderson, J.~(1967). {\it Principles of Relativity Physics}. Academic Press, New York.\\
\\
Anninos, D., Hartman, T., Strominger, A.~(2011). ``Higher Spin Realization of the dS/CFT Correspondence,''
  arXiv:1108.5735 [hep-th].\\
\\
Anninos, D., Denef, F., Harlow, D.~(2013). ``Wave function of Vasiliev’s universe: A few slices thereof,''
{\it Physical Review D}, {\bf 88} 8,  084049
  [arXiv:1207.5517 [hep-th]].\\
\\
Anninos, D., Mahajan, R., Radicevi\'c, R, Shaghoulian, E.~(2015). ``Chern-Simons-Ghost Theories and de Sitter Space,''
  {\it Journal of High Energy Physics} {\bf 1501} 074
  [arXiv:1405.1424 [hep-th]].\\
\\
Baggio, M., de Boer, J., Jottar, J. I., Mayerson, D. R. (2013). ``Conformal Symmetry for Black Holes in Four Dimensions and Irrelevant Deformations,''
  {\it Journal of High Energy Physics} {\bf 1304}, 084 
  [arXiv:1210.7695 [hep-th]].\\
\\
Balasubramanian, V.~and P.~Kraus (1999). ``A Stress tensor for Anti-de Sitter gravity,''
  {\it Communications in Mathematical Physics} {\bf 208} 413-428 
  [hep-th/9902121].\\
\\
Bedau, M. and Humphreys, P. (eds.) (2008), {\it Emergence: contemporary readings in 
philosophy and science}, MIT Press: Bradford Books. \\
\\
Bekenstein, J. D. (1973), ``Black holes and entropy,'' {\it Physical Review} \textbf{D7} 2333-2346. \\
\\
Bell, J. S. (1990), ``Against measurement'', in the second edition (2004) of his {\em Speakable and Unspeakable in Quantum Mechanics}, Cambridge: University Press.  \\
\\
Belot, G.~(2011). ``Background-Independence,''  {\it General Relativity and Gravitation}  {\bf 43} 2865-2884 
  [arXiv:1106.0920 [gr-qc]].\\
\\
Bena, I., El-Showk, S., Vercnocke, B. (2013), ``Black Holes in String Theory,'', in: Bellucci, S.~(ed.). {\it Black Holes in Supergravity}.  Springer Proceedings in Physics  {\bf 144}, 59-178 .\\
\\
Bena, I., Warner, N. P. (2008), ``Black holes, black rings and their microstates,''
  {\it Lecture Notes in Physics},  {\bf 755}, 1-92 
  [hep-th/0701216].\\
\\
Bigazzi, F., Cotrone, A. L., Mas, J., Mayerson, D., Tarrio, J. (2012). ``Holographic Duals of Quark Gluon Plasmas with Unquenched Flavors,''
  {\it Communications in Theoretical Physics}  {\bf 57}, 364-386 
  [arXiv:1110.1744 [hep-th]].\\
  \\
Bouatta, N. and Butterfield J. (2015), `On Emergence in Gauge Theories at the `t Hooft Limit', {\it European Journal for
Philosophy of Science} {\bf 5}, 55-87; http://arxiv.org/abs/1208.4986; http://philsci-archive.pitt.edu/9288 \\
\\
Bouatta, N. and Butterfield J. (2015a), ``Renormalization for philosophers'', in {\it Metaphysics in Contemporary Physics}, ed. T. Bigaj and C. Wuethrich (Poznan Studies in Philosophy of Science, vol 103; Rodopi, 2015), pp. 95-144.
http://arxiv.org/abs/1406.4532; http://philsci-archive.pitt.edu/10763/ \\
\\
Brown, J.D., York, Jr., J.W.~(1993). ``Quasilocal energy and conserved charges derived from the gravitational action,''
  {\it Physical Review D}, {\bf 47} 1407-1419 
  [gr-qc/9209012].\\
\\
Buchel, A., Liu, J. T., ``Universality of the shear viscosity in supergravity,''
  {\it Physical Review Letters}  {\bf 93}, 090602 (2004)
  [hep-th/0311175].\\
  \\
Butterfield J. (2011), ``Emergence, Reduction and Supervenience: a Varied Landscape'', {\it Foundations of Physics}, {\bf 41},  920-960. At Springerlink: doi:10.1007/s10701-011-9549-0; http://arxiv.org/abs/1106.0704: and at: http://philsci-archive.pitt.edu/5549/ \\
\\
Butterfield J. (2011a), ``Less is Different: Emergence and Reduction Reconciled'', {\it Foundations of Physics}, {\bf 41},1065-1135. At: Springerlink (DOI 10.1007/s10701-010-9516-1); http://arxiv.org/abs/1106.0702;  
and at: http://philsci-archive.pitt.edu/8355/ \\
\\
Butterfield J. (2014), ``Reduction, Emergence and Renormalization'', {\it The Journal of Philosophy} {\bf 111}, 5-49, At : http://arxiv.org/abs/1406.4354; http://philsci-archive.pitt.edu/10762/ \\
\\
Cardy, J. (2008), ``Conformal Field Theory and Statistical Mechanics,'' lectures given at the Summer School on
Exact methods in low-dimensional statistical physics and
quantum computing, les Houches, July 2008 [arXiv:0807.3472 [cond-mat.stat-mech]]\\
\\
Castellani, E. (2010). ``Dualities and intertheoretic relations", pp. 9-19 in: Suarez, M., M. Dorato and M. Red\'{e}i (eds.). \emph{EPSA Philosophical Issues in the Sciences}. Dordrecht: Springer.\\
\\
Chamblin, A., Lambert, N.D.~(2001). ``de Sitter space from M theory,''
  {\it Physics Letters B}, {\bf 508} 369-374 
  [hep-th/0102159].\\
\\
Chandler, D. (1987), {\it Introduction to Modern Statistical Mechanics}, New York, Oxford: Oxford University Press.\\
\\
Codesido, S., Grassi, S., and Mari\~no, M.~(2015). ``Exact results in $ \mathcal{N}=8 $ Chern-Simons-matter theories and quantum geometry,''
{\it Journal of High Energy Physics} {\bf 1507} 011
  [arXiv:1409.1799 [hep-th]].\\
\\
Compere, G. (2012), ``The Kerr/CFT correspondence and its extensions: a comprehensive review,''
  {\it Living Reviews in Relativity}  {\bf 15}, 11 
  [arXiv:1203.3561 [hep-th]].\\
\\
Cremonini, S. (2011), ``The Shear Viscosity to Entropy Ratio: A Status Report,''
  {\it Modern Physics Letters} B {\bf 25}, 1867-1888
  [arXiv:1108.0677 [hep-th]].\\
\\
de Boer, J., Verlinde, E., Verlinde, H. (2000), ``On the holographic renormalization group,''
  {\it Journal of High Energy Physics} {\bf 0008}, 003 (2000)
  [hep-th/9912012].\\
 \\
 Dawid, R. (2007), ``Scientific realism in the age of string theory'', {\it Physics and Philosophy}, {\bf 011}, 2007; PhilSci 3584; and http://physphil.uni-dortmund.de \\
 \\
De Haro, S. (2016), ``Dualities and emergent gravity: Gauge/gravity duality'', PhilSci 11666; http://arxiv.org/abs/1501.06162; forthcoming in {\em Studies in the History and Philosophy of Modern Physics}. doi:10.1016/j.shpsb.2015.08.004. \\
\\
De Haro, S. (2016a). ``Invisibility of Diffeomorphisms''. {\it Foundations of Physics}, submitted.\\
\\
De Haro, S.~(2016b). ``Reply to James Read on Background-Independence'', published in {\it Beyond Spacetime}: https://beyondspacetime.net/2016/06/15/reply-to-read.\\
\\
De Haro, S., Skenderis, K., and Solodukhin, S. (2001). ``Holographic reconstruction of spacetime and renormalization in the AdS/CFT correspondence", \emph{Communications in Mathematical Physics}, 217(3), 595-622. [hep-th/0002230].\\
\\
De Haro, S., Teh, N. and Butterfield J. (2016), ``Comparing dualities and gauge symmetries'', forthcoming in {\em Studies in the History and Philosophy of Modern Physics}. doi:10.1016/j.shpsb.2016.03.001.\\
\\
Deser, S.~, Schwimmer, A.~(1993).  ``Geometric classification of conformal anomalies in arbitrary dimensions,''
  {\it Physics Letters B}, {\bf 309}  279-284
  [hep-th/9302047].\\
\\
Dieks, D., Dongen, J. van, Haro, S. de, (2015), ``Emergence in Holographic Scenarios for Gravity'', PhilSci 11271, arXiv:1501.04278 [hep-th]. {\it Studies in History and Philosophy of Modern Physics,} 52(B), 203-216. doi:10.1016/j.shpsb.2015.07.007.\\
\\
Di Francesco, P.~, Mathieu, P.~, S\'en\'echal (1996). {\it Conformal Field Theory}. Springer-Verlag, New York.\\
\\
Farkas, H.~M.~, Kra, I.~(1980). ``Riemann Surfaces'', Springer-Verlag, New York.\\
\\
Gaberdiel, M. R. (1999), ``An introduction to conformal field theory,'' {\it Reports on Progress in Physics} {\bf 63}, 607-667  (2000)
  [hep-th/9910156].\\
\\
Ginsparg, P. H. (1988), ``Applied Conformal Field Theory,''  lectures given at Les Houches summer session, June 28--Aug. 5, 1988 [hep-th/9108028], in: Brezin, E.~and Zinn-Justin, J.~(1990). {\it Fields, Strings, Critical Phenomena}. North-Holland: Amsterdam. \\
\\
Giulini, D.~(2007). ``Some remarks on the notions of general covariance and background independence,''  {\it Lecture Notes in Physics}  {\bf 721} 105-120
  [gr-qc/0603087].\\
\\
Green, M.~B.~(1999). ``Interconnections between type II superstrings, M theory and N=4 supersymmetric Yang-Mills,''
  {\it Lecture Notes in Physics}, {\bf 525} 22
  [hep-th/9903124].\\
\\
Green, M. B., Schwarz, J. H., Witten, E. (1987), {\it Superstring Theory. Vol. 1: Introduction} and {\it Superstring Theory. Vol. 2: Loop Amplitudes, Anomalies and Phenomenology,} Cambridge, UK: Cambridge Monographs in Mathematical Physics. \\
\\
 Gross,  D.~J.~(1988). ``High-Energy Symmetries of String Theory,''
  {\it Physical Review Letters}  {\bf 60}  1229.\\
\\
Gubser, S.~S., Klebanov, I.~R., Polyakov, A.~M.~(1998). ``Gauge theory correlators from noncritical string theory,''
{\it Physics Letters B}, {\bf 428} 105-114 
  [hep-th/9802109].\\
\\
Hartnoll, S. A., Herzog, C. P., Horowitz, G. T. (2008), ``Building a Holographic Superconductor,'' {\it Physical Review Letters} \textbf{101} 031601.\\
\\
Hawking, S. W. (1974), ``Black hole explosions,'' {\it Nature} \textbf{248}, 30-31.\\
\\
Hawking, S. W. (1975), ``Particle Creation by Black Holes,'' {\it Communications in Mathematical Physics} \textbf{43}  199-200. \\
\\
Heemskerk, I.~and Polchinski, J.~(2011). ``Holographic and Wilsonian Renormalization Groups,''
  {\it Journal of High Energy Physics} {\bf 1106} 031
  [arXiv:1010.1264 [hep-th]].\\
\\
Henningson, M.~, Skenderis, K.~(1998). ``The Holographic Weyl anomaly,''  {\it Journal of High Energy Physics} {\bf 9807}  023
  [hep-th/9806087].\\
\\
Horowitz, G.~and Polchinski, J.~(2006), `Gauge/gravity duality', in {\em Towards quantum gravity?}, ed. Daniele Oriti, Cambridge University Press; arXiv:gr-qc/0602037\\
\\
  Huggett, N. (2016), ``Target space $\neq$ space'', PhilSci 11638; forthcoming in {\em Studies in the History and Philosophy of Modern Physics}. doi:10.1016/j.shpsb.2015.08.007.\\
\\
Hull, C.~M.~(2001). ``De Sitter space in supergravity and M theory,''
  {\it Journal of High Energy Physics} {\bf 0111} 012
  [hep-th/0109213].\\
\\
Kachru, S., Kallosh, R., Linde, A.D., Trivedi, S.P.~(2003). ``De Sitter vacua in string theory,''
  {\it Physical Review D}, {\bf 68} 046005
  [hep-th/0301240].\\
\\
Kallosh, R.~(2001). ``N=2 supersymmetry and de Sitter space,''
  hep-th/0109168.\\
\\
Karch, A., Katz, E. (2002). ``Adding flavor to AdS/CFT,''
  {\it Journal of High Energy Physics} {\bf 0206}, 043 
  [hep-th/0205236].\\
  \\
Ketov, S. V. (1995), {\it Conformal Field Theory}, World Scientific Publishing, Singapore. \\
\\
Kovtun, P., Son, D. T., Starinets, A. O.,``Viscosity in strongly interacting quantum field theories from black hole physics,''
  {\it Physical Review Letters}  {\bf 94}, 111601 
  [hep-th/0405231].\\
  \\
Kretschmann, E. (1917), ``{\"{U}}ber den physikalischen Sinn der Relativit{\"{a}}tspostulate.
{\it Annalen der Physik}, {\bf 53}, 575-614.\\
\\
Maldacena, J. (1998). ``The large \emph{N} limit of superconformal field theories and supergravity'',  \emph{Advances in Theoretical and Mathematical Physics} 2, 231-252.
  [hep-th/9711200].\\
\\
Maldacena, J.~M.~(2003). ``Non-Gaussian features of primordial fluctuations in single field inflationary models,''
{\it Journal of High-Energy Physics}, {\bf 0305} 013
  [astro-ph/0210603].\\
\\
Maldacena, J. (2003). ``TASI Lectures 2003 on AdS/CFT'', arxiv: hep-th/0309246, in: {\it Progress in string theory. Proceedings, Summer School, TASI 2003, Boulder, USA}, June 2-27, 2003.
Hackensack, USA: World Scientific. \\
\\
Maldacena, J. (2004), ``Black Holes and Holography in String Theory'', in: B. Duplantier and V. Rivasseau (Eds): {\it Seminaire Poincare}, 61-67.\\
\\
Maldacena, J. (2004a), ``Quantum gravity as an ordinary gauge theory'', in J. Barrow, P. Davies and C. Harper ed.s, {\em Science and Ultimate Reality: quantum theory, cosmology and complexity}, Cambridge University Press, pp. 153-166. \\
\\
Maldacena, J.~M.~(2011). ``Einstein Gravity from Conformal Gravity,''
  arXiv:1105.5632 [hep-th]. \\
  \\
  Matsubara, K. (2013), ``Realism, underdetermination and string theory dualities'', {\it Synthese}, {\bf 190} (3): 471-489.\\
\\
Mathur, S. D. (2008), ``Fuzzballs and the information paradox: A Summary and conjectures,''
  arXiv:0810.4525 [hep-th].\\
\\
McGreevy, J. (2010), ``Holographic duality with a view to many-body physics''. {\em Advances in High Energy Physics} 723105 [arXiv:0909.0518 [hep-th]].\\
\\
Ng, G.~S.~and A.~Strominger (2013). ``State/Operator Correspondence in Higher-Spin dS/CFT,''
{\it  Classical and Quantum Gravity}  {\bf 30}  104002
  [arXiv:1204.1057 [hep-th]].\\
\\
Perlmutter, S.~(2011). ``Measuring the Acceleration of the Cosmic Expansion Using Supernovae'', Nobel Lecture, http://www.nobelprize.org/nobel\_prizes/physics/laureates/2011/perlmutter-lecture.html.\\
\\
Petersen, J. (1999). {\it International Journal of Modern Physics} A {\bf 14} 3597-3672 
  [hep-th/9902131].\\
\\
Polchinski, J. (1998), {\it String Theory. Vol 1.: An introduction to the bosonic string,} and {\it String Theory. Vol 2.: Superstring theory and beyond}, Cambridge, UK: Cambridge University Press.\\
\\
Policastro, G., Son, D. T., Starinets, A. O. (2001), ``The Shear viscosity of strongly coupled N=4 supersymmetric Yang-Mills plasma,''
  {\it Physical Review Letters}  {\bf 87}, 081601 
  [hep-th/0104066].\\
\\
Read, J.~(2016). ``Reply to Sebastian De Haro on Background Independence'', published in {\it Beyond Spacetime}: https://beyondspacetime.net/2016/06/15/reply-to-read.\\
\\
Rickles, D.~(2011). ``A philosopher looks at string dualities'',  \emph{Studies in History and Philosophy of Science Part B: Studies in History and Philosophy of Modern Physics}, 42(1), 54-67.\\
\\
Rickles, D.~(2012). ``AdS/CFT duality and the emergence of spacetime'', \emph{Studies in History and Philosophy of Science Part B: Studies in History and Philosophy of Modern Physics}, 44(3), 312-320.\\
\\
Rickles, D. (2016). ``Dual theories: `same but different' or different but same'?'', forthcoming in {\em Studies in the History and Philosophy of Modern Physics}. doi:10.1016/j.shpsb.2015.09.005.\\
\\
Sakai, T., Sugimoto, S. (2005). ``Low energy hadron physics in holographic QCD,''
{\it Progress in Theoretical Physics}  {\bf 113}, 843-882
  [hep-th/0412141].\\
  \\
Shuryak, E. V. (2004), ``What RHIC experiments and theory tell us about properties of quark-gluon plasma?,''
 {\it Nuclear Physics} A {\bf 750}  (2005) 64-83
  [hep-ph/0405066].\\
\\
Skenderis, K. (2002), `Lecture notes on holographic renormalization', {\em Classical and Quantum Gravity} {\bf 19}, pp. 5849-5876.\\
\\
Skenderis, K.~and B.~C.~van Rees (2009).   ``Real-time gauge/gravity duality: Prescription, Renormalization and Examples,''
  {\it Journal of High Energy Physics} {\bf 0905}  085
  [arXiv:0812.2909 [hep-th]].\\
\\
Son, D. T., Starinets, A. O. (2007), ``Viscosity, Black Holes, and Quantum Field Theory,'' {\it Annual Review of Nuclear and Particle Science} {\bf 57}, 95-118 
  [arXiv:0704.0240 [hep-th]].\\
\\
Spradlin, M., Strominger, A., Volovich, A.~(2001). ``Les Houches lectures on de Sitter space,''
  hep-th/0110007, in: {\it Unity from duality: Gravity, gauge theory and strings. Proceedings, NATO Advanced Study Institute, Euro Summer School, 76th session, Les Houches, France, July 30-August 31, 2001}, C. Bachas, A. Bilal, M. Douglas, N. Nekrasov, F. David (Eds.), Les Ulis: (2002). \\
\\
Strominger, A.~(2001). ``The dS / CFT correspondence,''
  {\it Journal of High-Energy Physics}, {\bf 0110} 034
  [hep-th/0106113].\\
\\
Strominger, A.~(2001). ``Inflation and the dS/CFT correspondence,''
  {\it Journal of High Energy Physics} {\bf 0111}  049
  [hep-th/0110087].\\
\\
Teh, N.J.~(2013). ``Holography and emergence'', \emph{Studies in History and Philosophy of Science Part B: Studies in History and Philosophy of Modern Physics}, 44(3), 300-311.\\
\\
't Hooft, G. (1974), ``A Planar Diagram Theory for Strong Interactions,''
  {\it Nuclear Physics} B {\bf 72}, 461-473.\\
\\
't Hooft, G. (1990), ``The black hole interpretation of string theory,''
  {\it Nuclear Physics} B {\bf 335}, 138-154 .\\
\\
't Hooft, G.~(1993). ``Dimensional reduction in quantum gravity'', in: Ali, A., J. Ellis and S. Randjbar-Daemi (Eds.), \emph{Salamfestschrift}. Singapore: World Scientific. [gr-qc/9310026].\\
\\
't Hooft, G.~(2013). ``On the Foundations of Superstring Theory'', {\it Foundations of Physics}, 43 (1), 46-53\\ 
\\
Tong, D. (2009). ``Lectures on String Theory'', arXiv:0908.0333 [hep-th].\\
\\
Townsend, P. (1997), ``Black Holes: Lecture Notes,'' arXiv:gr-qc/9707012 [gr-qc].\\
\\
Vasiliev, M.~A.~(1990). ``Consistent equation for interacting gauge fields of all spins in (3+1)-dimensions,''
 {\it Physics Letters} B {\bf 243} 378-382 .\\
\\
Vasiliev, M.~A.~(1999). ``Higher spin gauge theories: Star product and AdS space,''
  In *Shifman, M.A. (ed.): The many faces of the superworld* 533-610
  [hep-th/9910096].\\
\\
 Vasiliev,  M.~A.~(2003). ``Higher spin gauge theories in various dimensions,''
{\it  Fortschritte der Physik} {\bf 52} (2004) 702-717
   [{\it Progress of Physics} (2003) 003 Johns Hopkins Workshop on Current Problems in Particle Theory : Symmetries and Mysteries of M-Theory]
  [hep-th/0401177].\\
\\
Verlinde, E. (2011). ``On the origin of gravity and the laws of Newton'', {\it Journal of
High Energy Physics,} 029. [arXiv:1001.0785 [hep-th]].\\
\\
Witten, E. (1998a). ``Anti-de Sitter space and holography,''
 {\it Advances in Theoretical and Mathematical Physics} {\bf 2}, 253 
  [hep-th/9802150].\\
\\
Witten, E. (1998b). ``Anti-de Sitter space, thermal phase transition, and confinement in gauge theories,''  
{\it Advances in Theoretical and Mathematical Physics}  {\bf 2}, 505-532 
 [hep-th/9803131].\\
 \\
 Witten, E.~(2001). ``Quantum gravity in de Sitter space,''
  hep-th/0106109. Strings 2001: International Conference 5-10 Jan 2001. Mumbai, India. \\
\\
Zwiebach, B. (2009). {\it A first course in string theory}, Cambridge, UK: Cambridge University Press.

\end{document}